\documentclass[a4paper,12pt]{article}

\usepackage{epsfig}
\usepackage{epstopdf}
\usepackage{a4}
\usepackage{latexsym, amssymb} 
\usepackage{ctable}
\usepackage{threeparttable}
\usepackage{footnote}
\usepackage{amsmath}
\usepackage{url}
\usepackage{graphicx}
\usepackage{float}
\usepackage{soul}
\usepackage{xcolor,cancel}
\newcommand{\lapprox}{%
\mathrel{%
\setbox0=\hbox{$<$}
\raise0.6ex\copy0\kern-\wd0
\lower0.65ex\hbox{$\sim$}
}}
\newcommand{\gapprox}{%
\mathrel{%
\setbox0=\hbox{$>$}
\raise0.6ex\copy0\kern-\wd0
\lower0.65ex\hbox{$\sim$}
}}

\newcommand{\ba}{\begin{array}}
\newcommand{\ea}{\end{array}}
\newcommand{\bd}{\begin{displaymath}}
\newcommand{\ed}{\end{displaymath}}
\newcommand{\be}{\begin{equation}}
\newcommand{\ee}{\end{equation}}
\newcommand{\bea}{\begin{eqnarray}}
\newcommand{\eea}{\end{eqnarray}}

\def\q2 {q^2}

\def\bt{\begin{table}}
\def\et{\end{table}}

\catcode`@=11 
\def \gsim{\mathrel{\mathpalette\@versim>}}
\def \lsim{\mathrel{\mathpalette\@versim<}}
\def \@versim#1#2{\lower0.4ex\vbox{\baselineskip\z@skip\lineskip\z@skip
     \lineskiplimit\z@\ialign{$\m@th#1\hfil##\hfil$%
     \crcr#2\crcr\sim\crcr}}}
\catcode`@=12 

\begin{document}

\begin{flushright}
{\small 
IPMU13-0160\\
RECAPP-HRI-2013-018}
\end{flushright}

\begin{center}

{\large\bf Higher dimensional operators and LHC Higgs data :\\ the role of modified kinematics}\\[15mm]
Shankha Banerjee\footnote{E-mail: shankha@hri.res.in},
Satyanarayan Mukhopadhyay\footnote{E-mail: satya.mukho@ipmu.jp}
and Biswarup Mukhopadhyaya\footnote{E-mail: biswarup@hri.res.in}\\ \bigskip
{\em $^{1,3}$Regional Centre for Accelerator-based Particle Physics \\
     Harish-Chandra Research Institute\\
Chhatnag Road, Jhusi, Allahabad - 211 019, India\\ \bigskip
$^2$Kavli IPMU (WPI), The University of Tokyo,\\
Kashiwa, Chiba 277-8583, Japan.}
\\[20mm] 
\end{center}

\begin{abstract} 

The inclusion of higher-dimensional gauge invariant operators induces
new Lorentz structures in Higgs couplings with electroweak gauge boson
pairs. This in principle affects the kinematics of Higgs production
and decay, thereby modifying the efficiencies of the experimental cuts
compared to what simulations based on the standard model interactions
yield. Taking some sample cases, we perform a rigorous analysis of how
the efficiencies differ for various strengths of the additional
operator vis-a-vis the standard model interactions, scanning over the
values of both of them. While the response to cuts can be markedly
different in some regions, we find that the sensitivity to new
operator structures is relatively limited, so long as we remain 
confined to the 2$\sigma$ regions around the best fit signal
strengths measured at the Large Hadron Collider. We also show modifications to certain
kinematical distributions including the new operators in the diphoton 
final state.

\end{abstract}

\vskip 1 true cm

\newpage
\setcounter{footnote}{0}

\def\baselinestretch{1.5}
\section{Introduction}
After the discovery of a new boson with a mass around 125 GeV at the Large Hadron Collider (LHC)~\cite{ATLAS,CMS}, there have been numerous studies attempting to pin down its properties, 
namely, its spin-parity and its couplings to standard model (SM) particles~\cite{before-discovery,higgs-pheno,higgs-pheno2,atlas2,atlas3,cms2}. The bosonic decay modes of this particle have been analyzed with greater precision 
than the fermionic modes by both the ATLAS and CMS collaborations, since the latter requires much more statistics and possibly the application of new search strategies. The signal-strengths 
reported in various channels by the experiments are broadly consistent with the SM predictions within about two standard deviations, and a preliminary analysis of spin-parity using the 
$ZZ^*\rightarrow 4\ell$ channel suggests that a CP-even scalar hypothesis is favoured over other alternatives~\cite{atlas-cp}. Therefore, the accumulating evidence is in favour of an SM-like 
($J^{PC}=0^{++}$) Higgs boson, and we are going to assume so henceforth. Global fits of the Higgs data have been used by both experimentalists and theorists to derive bounds on possible 
deviations from the SM. Such deviations in the Higgs couplings can be parametrized either by including a multiplicative (or additive) constant to the SM coupling, or by including new Lorentz 
structures not present in the renormalizable SM Lagrangian. In the framework of the SM as an effective field theory valid below a cut-off scale $\Lambda$, higher-dimensional operators 
involving the SM fields and invariant under the SM gauge group can be used to capture possible new physics effects.  A complete list of such operators has been written down in 
Ref.~\cite{Buchmueller}, while a minimal basis has been obtained rather recently in Ref.~\cite{min-basis}. Such an approach is valid as long as there is no new light degree of freedom coupled 
to the SM  sector below the scale $\Lambda$. Null results in LHC searches for new particles provide some motivation for this approach, although the presence of new particles charged under the 
SM gauge group is still viable even with masses around the electroweak symmetry breaking scale. Henceforth, we assume that there is no such new state, and work with the SM Lagrangian 
supplemented with dimension-6 operators involving SM gauge bosons and the SM Higgs doublet. As is well-known, although there exists one possible dimension-5 operator, it plays a role only in 
the generation of neutrino masses.  

Electroweak precision measurements constrain the overall strength of the operators involving SM electroweak gauge bosons~\cite{Hagiwara,Garcia}. However, such constraints come from one-loop 
contributions of these operators to the self-energy diagrams of the gauge-bosons, parametrized in terms of the so-called oblique corrections~\cite{Peskin,Altarelli}. In contrast, the Higgs 
boson couplings to W, Z or photon pairs can be affected at the tree level itself by a class of such operators, and therefore, it is possible to impose stronger constraints on their 
co-efficients using the already accumulated LHC Higgs data. This fact has been observed in several studies performing global fits to the Higgs data, and deriving limits on the operator 
co-efficients~\cite{Sanz,Garcia2,Falkowski}. However, in most cases, these studies make an important assumption, namely, that the efficiencies of experimental cuts used for various final 
states are the same as the corresponding efficiencies in the SM case. To understand where the efficiency of experimental cuts enter the global fits, let us 
recall that the global fits are performed by comparing the experimentally observed signal strength ($\hat{\mu}_{X \bar{X}}$) in a channel $X\bar{X}$  with the corresponding signal strength 
predicted by a  particular framework beyond the standard model (BSM),  $\mu_{X \bar{X}}$, which is defined as
\begin{equation}
\mu_{X \bar{X}} = \frac{\left[\sigma(pp \rightarrow H)\times {\rm BR} (H \rightarrow X \bar{X}) \times \epsilon_{X \bar{X}}\right]_{\rm BSM}}{\left[\sigma(pp \rightarrow H)\times {\rm BR} (H \rightarrow X \bar{X}) \times \epsilon_{X \bar{X}}\right]_{\rm SM}},
\label{mu}
\end{equation}
where, $\epsilon_{X \bar{X}}$ denotes the efficiency of the experimental cuts applied to select a particular final state. Although the assumption that 
$(\epsilon_{X \bar{X}})_{\rm BSM}=(\epsilon_{X \bar{X}})_{\rm SM}$ can be justified if the Higgs couplings only receive a multiplicative modification to the SM one, it is not {\em a priori} 
clear whether such an assumption holds after the inclusion of dimension-6 operators. This is because these operators bring in new Lorentz structures to the Higgs-gauge boson couplings, 
which in turn modify the distributions of kinematic variables on which these cuts are imposed. Some of these distributions have been used in earlier studies, with special emphasis on the 
spin-parity determination of the newly discovered particle~\cite{Plehn,Mellado,spin-cp}. In this paper, we assume $J^{PC}=0^{++}$ for this particle, and investigate how additional interaction terms 
with gauge boson pairs, gauge invariant and of higher dimension, affect Higgs phenomenology. With this in view, we subject the contributions of the additional operators to the cuts used on 
specific final states. Thus we demonstrate through rigorous Monte Carlo simulations how much the efficiencies can get modified, and to what extent they alter the bounds on the operator co-efficients. 
We use LHC Higgs search studies in the $WW^*$ and $\gamma \gamma$ channels as examples, implement the cuts used by the ATLAS collaboration in our toy detector simulation, and determine 
the modified efficiencies for two such dimension-6 operators\footnote{For an analysis of the modified efficiencies in the $Z Z^{*}$ channel, we refer the readers to Ref.~\cite{Matchev}.}. 
We also simultaneously allow the modification of the SM coupling to the weak gauge bosons by a multiplicative constant, keeping 
the custodial $SU(2)$ symmetry intact. It should be mentioned that generically more than one higher-dimensional operator can be 
present in the effective low-energy theory with different coupling strengths. In that sense, our study with one operator 
considered at a time is illustrative, and focuses on the important effect of new Lorentz structures in the cut efficiencies. Moreover, the method developed 
here is of general utility in studying all possible higher-dimensional operators.

This paper is organized as follows. In section~\ref{sec-2}, we describe the higher dimensional operators considered, the modified Higgs-gauge boson couplings that they lead to, and constraints 
on them from electroweak precision tests. In section~\ref{sec-3}, we describe the set-up of our Monte-Carlo simulation, including its validation against the ATLAS Higgs search studies in the 
$WW^*$ channel. The modified decay widths, cross-sections and efficiencies are presented in section~\ref{sec-4}, including simple parametrizations of each of these. Section~\ref{sec-5} is 
devoted to the re-evaluation of constraints on these operators using a fit to the Higgs data in bosonic channels, while in section~\ref{sec-6} we study the modified efficiencies in the associated 
production of Higgs. In section~\ref{sec-7} we show the modifications to certain kinematic 
observables in presence of the new operators. We summarize our findings in section~\ref{sec-8}.

\section{Dimension-6 operators and electroweak precision constraints}
\label{sec-2}
To illustrate the modification of experimental cut-efficiencies on including new Lorentz structures in Higgs-gauge boson interactions (henceforth called $HVV$ interactions) we take the 
following dimension-6 operators as examples:
\begin{align}
\mathcal{O}_{WW} &= {\Phi}^{\dagger}\hat{W}_{\mu \nu} \hat{W}^{\mu \nu} \Phi \nonumber \\
\mathcal{O}_{BB} &= {\Phi}^{\dagger}\hat{B}_{\mu \nu} \hat{B}^{\mu \nu} \Phi,
\label{ops}
\end{align}
where, the field strength tensors for the $SU(2)_{\rm L}$ and $U(1)_Y$ gauge groups are 
\begin{align}
\hat{W}_{\mu \nu} &= i \frac{g}{2}\sigma^{a} \left(\partial_{\mu}W_{\nu}^{a} - \partial_{\nu}W_{\mu}^{a} - g \epsilon^{abc} W_{\mu}^{b} W_{\nu}^{c} \right)\nonumber \\
\hat{B}_{\mu \nu} &= i \frac{g^{\prime}}{2} \left(\partial_{\mu} B_{\nu} - \partial_{\nu} B_{\mu}\right).
\end{align}
Here, $\Phi$ denotes the SM Higgs doublet, $\sigma^a$ are the Pauli matrices, and $g$, $g^{\prime}$ are the $SU(2)_{\rm L}$, $U(1)_Y$ gauge couplings respectively.
As mentioned in the introduction, the SM $HWW$ and $HZZ$ couplings can also get modified by a multiplicative factor in a general setting, where, for example, the SM Higgs doublet is part of a 
larger scalar sector. We therefore include the possibility of having the $HWW$ and $HZZ$ couplings modified by the same factor $\beta$, assuming custodial invariance. Since generically, in 
presence of an arbitrary number of extra scalar singlets or doublets, $\beta\leq1$~\cite{Wudka}, we scan the range $0<\beta\leq1$ in our analysis. One of the goals of this study is to determine to 
what extent $\beta$ can be different from its SM value of $1$, while including new dimension-6 operators. Taking these two modifications into account, the Lagrangian in the Higgs sector 
becomes

\begin{equation}
\mathcal{L} = \mathcal{L}_{SM} (\beta) + \frac{f_{WW}}{\Lambda^2}\mathcal{O}_{WW} + \frac{f_{BB}}{\Lambda^2}\mathcal{O}_{BB}~,
\label{Lag}
\end{equation}
where, the operators $\mathcal{O}_{WW}$ and $\mathcal{O}_{BB}$ are given by equation~\ref{ops}, and  the $\beta$ dependence of the SM part comes from the following terms
\begin{equation}
\mathcal{L}_{SM} (\beta) \supset \beta \left(\frac{2 m_W^2}{v} H W_{\mu}^+ W^{\mu -}+\frac{ m_Z^2}{v} H Z_{\mu} Z^{\mu }      \right).
\label{SM-beta}
\end{equation}
Here, $v$ denotes the vacuum expectation value of the Higgs field, and $H$ denotes the Higgs boson. The Higgs couplings in $\mathcal{L}_{SM}$ to fermions and gluons are not modified. 

Since the new Higgs couplings generated by $\mathcal{O}_{WW}$ and $\mathcal{O}_{BB}$, in particular the new Lorentz structures, are crucial to our discussion, we note down the additional 
$HVV$ interactions generated by the operators in equation~\ref{ops}~\cite{Hagiwara,Garcia}:
\begin{equation}
\mathcal{L}_{HVV} \supset  g_{H WW}H W_{\mu \nu}^{+} W^{\mu \nu}_{-} +
                           g_{H ZZ}H Z_{\mu \nu} Z^{\mu \nu} +
                           g_{H \gamma \gamma}H A_{\mu \nu} A^{\mu \nu} + 
                           g_{H Z \gamma}H A_{\mu \nu} Z^{\mu \nu},
\label{Interactions}
\end{equation}
where, $V_{\mu \nu} = \partial_{\mu}V_{\nu} - \partial_{\nu}V_{\mu}$, for $V = \lbrace\gamma, W^{\pm}, Z\rbrace$, and,
\begin{align}
\label{coupling-change}
g_{H WW} &= -\left(\frac{g m_W}{\Lambda^2}\right)f_{WW}\nonumber \\
g_{H ZZ} &= -\left(\frac{g m_W}{\Lambda^2}\right)\frac{s^4 f_{BB} + c^4 f_{WW}}{2 c^2}\nonumber \\
g_{H \gamma \gamma} &= -\left(\frac{g m_W}{\Lambda^2}\right)\frac{s^2(f_{BB} + f_{WW})}{2} \nonumber \\
g_{H Z \gamma} &= \left(\frac{g m_W}{\Lambda^2}\right)\frac{s(s^2 f_{BB} - c^2 f_{WW})}{c}. 
\end{align}
We have used the shorthand $c=\cos \theta_W$ and $s=\sin \theta_W$, $\theta_W$ being the Weinberg angle. The interaction terms involving the derivatives of the gauge fields bring in momentum 
dependent vertices, which are responsible for the modified kinematics in Higgs boson production via weak-boson fusion and associated production with a $W$ or a $Z$, as well as the decay of 
the Higgs particle to electroweak gauge boson final states.  The kinematics is affected most when the new interactions appear in both the production and decay processes, an example of which we shall discuss 
in sections~\ref{sec-3} and \ref{sec-4}. We should remark here that in $g_{H \gamma \gamma}$, only the new tree-level terms generated due to $\mathcal{O}_{WW}$ and $\mathcal{O}_{BB}$ have 
been considered. There will be additional contributions coming from the $W$ boson loop (apart from the SM contribution modified by the inclusion of $\beta$, which we take into account), since 
now the $HWW$ coupling also involves momentum-dependent terms. However, on naive power-counting in the number of loops and derivatives, these contributions are sub-leading, and the new 
divergences arising from this loop diagram will be cancelled by the next higher-order terms in the derivative expansion\footnote{We thank Adam Falkowski for clarifying this point.}. With the 
current level of precision in the data, such terms can be safely neglected. Finally, though we have noted the new contribution to the $HZ\gamma$ vertex for completeness, there is no data in 
this channel so far, and therefore, the effect of the modification to this channel is sub-dominant.

The operators $\mathcal{O}_{WW}$ and $\mathcal{O}_{BB}$ contribute to the so-called Peskin-Takeuchi $STU$ parameters~\cite{Peskin,Altarelli}, and are therefore constrained by electroweak 
precision data~\cite{Hagiwara,Garcia}. Following Ref.~\cite{Garcia}, the bounds at $95 \%$ C.L., taking one operator at a time, are given by
\begin{align}
-24~{\rm TeV}^{-2}<\frac{f_{WW}}{\Lambda^2}<14~{\rm TeV}^{-2} \nonumber \\
-79~{\rm TeV}^{-2}<\frac{f_{BB}}{\Lambda^2}<47~{\rm TeV}^{-2}.
\end{align}
These bounds can change once we include the factor $\beta$ in equation~\ref{SM-beta}. However, as we shall see, the Higgs data puts much stronger constraints on these two operators compared 
to the precision data, and therefore, we do not consider modifications to the precision constraints in this study.

We note in passing that since the inclusion of the higher-dimensional operators in equation~\ref{Lag} modifies the $HWW$ and $HZZ$ vertices from their corresponding SM values, this will spoil the unitarity of $V_L V_L \rightarrow V_L V_L$ ($V=W,Z$) scattering amplitudes at high energies.  For the values of operator coefficients allowed by the current Higgs data, the violation of unitarity appears at energies of a few TeV, with the exact value depending upon the specific choice of operators~\cite{Unitarity}.  Since the higher-dimensional operators themselves arise from integrating out  heavy fields of mass $\mathcal{O}(\Lambda)$ (for weakly coupled ultra-violet completions), one expects in general that the presence of these new degrees of freedom in the UV completion will eventually restore the unitarity of the full theory at high energies.

\section{Simulation framework and validation using \\ $H\rightarrow WW^{*} + \geq 2j$ data}
\label{sec-3}
As noted in the previous section, we expect in general significant modifications to the kinematics in processes where the new Lorentz structures appear both in the production and decay 
vertices. An example of such a process is the production of Higgs boson via VBF and its subsequent decay to $WW^*$. The ATLAS collaboration has presented a detailed analysis of such 
a scenario in the $WW^* \rightarrow \ell^+ \nu \ell^- \bar{\nu}$ channel ($\ell=\lbrace e, \mu \rbrace$) in Ref.~\cite{ATLAS-WW} using the $8$ TeV, $ 20~ {\rm fb}^{-1}$ data. The response of 
SM-type interactions to all the individual cuts can be readily checked from this analysis, which thus provides a much needed calibration for the simulation with additional operators. Therefore, 
we use this channel to set-up and validate our Monte Carlo as well as our detector simulation code.

We have used {\tt FeynRules}~\cite{FR} to extract the Feynman rules from the Lagrangian in equation~\ref{Lag}, {\tt MadGraph-5}~\cite{MG5} to generate the parton level events, 
{\tt Pythia-6}~\cite{Pythia} for parton shower and hadronization, and our own detector simulation code for analyzing the hadron-level events. Jet formation and underlying events have been 
simulated within the {\tt Pythia} framework.  
\begin{table}[t]
\centering
\begin{tabular}{|l|c|c|}
\hline
Cut & ATLAS efficiency  & Our MC efficiency \\
\hline
$N_{b-jet}=0$ & 0.68-0.76 (0.72) & 0.74  \\
$p_{T}^{tot} < 45$ & 0.81-0.93 (0.87) & 0.88 \\
$Z \rightarrow \tau \tau$ veto & 0.86-1.00 (0.92) & 0.95 \\
$|\Delta y_{jj}| > 2.8$ & 0.45-0.51 (0.48) & 0.50 \\
$m_{jj} > 500$ & 0.61-0.64 (0.62) & 0.53 \\
No jets in $y$ gap & 0.82-0.86 (0.84) & 0.81 \\
Both $l$ in $y$ gap & 0.94-1.00 (0.97) & 0.95 \\
$m_{ll} < 60$ & 0.87-0.93 (0.90) & 0.95 \\
$| \Delta \phi_{ll} | < 1.8$ & 0.89-0.96 (0.93) & 0.92 \\
\hline
\end{tabular}
\caption{\small \sl Comparison of the efficiencies of experimental cuts on the signal cross-section in the $H \rightarrow WW^* \rightarrow\ell^+ \nu \ell^- \bar{\nu}$ channel, for the 
$N_{jet} \geq 2$ category, demanding different flavour leptons ($e^+\mu^-+\mu^+e^-$) in the final state. The signal cross-section here refers to the sum of VBF and VH processes. The ATLAS 
numbers have been taken from Ref.~\cite{ATLAS-WW}, for which we show the $1\sigma$ range (the central value is written within brackets).}
\label{effOF}
\end{table}
\begin{table}[htb!]
\centering
\begin{tabular}{|l|c|c|}
\hline
Cut & ATLAS efficiency  & Our MC efficiency\\
\hline
$N_{b-jet}=0$ & 0.69-0.77 (0.73) & 0.73 \\
$p_{T}^{tot} < 45$ & 0.84-0.95 (0.89) & 0.87 \\
$|\Delta y_{jj}| > 2.8$ & 0.45-0.50 (0.48) & 0.50 \\
$m_{jj} > 500$ & 0.65-0.71 (0.68) & 0.57 \\
No jets in $y$ gap & 0.82-0.89 (0.85) & 0.81 \\
Both $l$ in $y$ gap & 0.92-1.00 (0.96) & 0.93 \\
$m_{ll} < 60$ & 0.85-0.93 (0.89) & 0.94 \\
$| \Delta \phi_{ll} | < 1.8$ & 0.88-0.97 (0.92) & 0.90 \\
\hline
\end{tabular}
\caption{\small \sl Same as table~\ref{effOF}, for same-flavour leptons in the final state ($e^+e^-+\mu^+\mu^-$).}
\label{effSF}
\end{table}

In the study of Higgs boson decaying to $WW^*$, followed by the semi-leptonic decay of the $W$'s, the ATLAS collaboration has considered three categories, namely, the production of Higgs in 
association with $0,1,$ and $\geq 2$ jets. In this part of the study, we consider only the $\geq 2$-jet category. According to Ref.~\cite{ATLAS-WW}, vector-boson fusion (VBF) and 
associated production with $W$ or $Z$ (called VH) are considered as signals in this category. For the $WW^{*}$ final state, the VBF channel is picked out in the ATLAS analysis, by requiring a 
high invariant mass for the two leading jets in the forward region. The gluon-fusion production of Higgs is considered as a background. For a detailed description of the experimental cuts 
used, we refer the reader to Ref.~\cite{ATLAS-WW}. To validate our Monte Carlo (MC) simulation, in tables~\ref{effOF} and \ref{effSF}, we compare the efficiencies of each of the experimental 
cuts obtained by our MC in the SM case, with the numbers reported by ATLAS, in the same and opposite flavour dilepton sub-categories respectively. As we can see from this comparison, our 
simulations agree with the ATLAS simulation to within $5 \%$ for all cuts except the one on $m_{jj}$, for which the difference is $\sim 15 \%$. Our simulation shows a lower efficiency for 
this cut compared to ATLAS, and a possible reason for this is our inadequate modelling of detector effects for jets. Since the purpose of this part of the study is the overall validation of 
our MC, and in the subsequent sections we concentrate on the modification of efficiencies after including the dimension-6 operators within our own MC set-up, this difference is not expected 
to alter our main conclusions. 

\section{Modified efficiencies and signal strengths}
\label{sec-4}
After the validation of our MC simulation framework in the previous section, we are now in a position to determine the modified cut-efficiencies ($(\epsilon_{X \bar{X}})_{\rm BSM}$ in 
equation~\ref{mu}) and signal strengths $\mu_{X \bar{X}}$ using the Lagrangian in equation~\ref{Lag}. We first do so in the $WW^*$ channel for the $\geq 2$-jet category considered in 
section~\ref{sec-3}, by including only the operator $\mathcal{O}_{WW}$, where we expect the maximum modification. The efficiency then is a function of the parameters $\beta$ and $f_{WW}$, and 
is given by
\begin{equation}
\epsilon_{WW^*+\geq 2{\rm -jets}} (\beta,f_{WW}) = \frac{\left[\sigma(pp \rightarrow H)_{\rm VBF+VH } \times {\rm BR} (H \rightarrow WW^*)  \right]_{\rm After ~Cuts}}{\left[\sigma(pp \rightarrow H)_{\rm VBF+VH } \times {\rm BR} (H \rightarrow WW^*)  \right]_{\rm Before ~Cuts}}.
\label{eff-def}
\end{equation}
The theoretically calculated efficiencies are assumed here to be independent of radiative corrections. We evaluate the cross-sections before and after cuts by scanning over the parameters 
$\beta$ and $f_{WW}$, and since they are found to be smooth functions of these parameters even after 
detector level simulations, we can parametrize them by simple polynomial functions of $\beta$ and $f_{WW}$. The Higgs boson partial decay widths in the 
$WW^*, ~ZZ^*,~\gamma \gamma$ and $Z\gamma$ channels are also functions of these two variables, while in the rest of the channels the partial widths are the same as in the SM. Since higher-order 
corrections are small in the aforementioned bosonic channels~\cite{Djouadi2}, we compute them at tree level, while for all other channels we have used the NNLO predictions from 
Ref.~\cite{LHCHWG} for a Higgs mass of  125 GeV. The tree-level partial widths (in GeV) in these channels are rather accurately parametrized by the following expressions :
\begin{align}
 \Gamma_{H \rightarrow W W^{*}} &= 8.61 \times 10^{-4}\beta^2 + 8.51 \times 10^{-6} \beta f_{WW} + 2.95 \times 10^{-8} f_{WW}^2 \nonumber \\ 
 \Gamma_{H \rightarrow Z Z^{*}} &=  9.28 \times 10^{-5}\beta^2 + 4.77 \times 10^{-7} \beta f_{WW} + 1.00 \times 10^{-9} f_{WW}^2 \nonumber \\
 \Gamma_{H\rightarrow \gamma \gamma} &= 8.59 \times10^{-7} - 8.04 \times 10^{-6} \beta - 4.36 \times 10^{-6} f_{WW} \nonumber \\
         & + 1.77 \times 10^{-5} \beta^{2} + 1.98 \times 10^{-5} \beta f_{WW} + 5.68 \times 10^{-6} f_{WW}^2 \nonumber \\
 \Gamma_{H\rightarrow Z \gamma} &= 3.75 \times 10^{-8} - 7.91 \times 10^{-7} \beta -5.65 \times 10^{-7} f_{WW} \nonumber \\
         & + 7.12 \times 10^{-6} \beta^2 + 1.06 \times 10^{-5} \beta f_{WW} + 3.82 \times 10^{-6} f_{WW}^2
 \label{partial}
\end{align}
For the above formulae and all subsequent ones involving $f_{WW}$, we have used a reference scale of $\Lambda=1{~\rm TeV}$, and for a different 
choice of the cut-off, the coefficients should be re-scaled according to the power of $f_{WW}$ involved. Adding all the contributions, the total Higgs boson width becomes
\begin{align}
\Gamma_{\rm tot} &= [3.07 -7.82 \times 10^{-3} \beta -4.37 \times 10^{-3} f_{WW} \nonumber \\ 
                 & + 0.97 \beta^2 + 3.67 \times 10^{-2} \beta f_{WW} + 8.76 \times 10^{-3} f_{WW}^2] \times 10^{-3}{\rm GeV}.
\label{total}
\end{align}
Similarly, the tree-level total cross-section for the VBF and VH processes at 8 TeV LHC, before the application of selection cuts,  can be expressed as follows
\begin{equation}
\sigma_{{p p \rightarrow H + 2{\rm -jets}}}{\rm (VBF + VH)} = \left(2.0432 \beta^2 - 0.0330 \beta f_{WW} + 0.0030 f_{WW}^2\right){\rm pb}.
\label{sigma}
\end{equation}
By performing a scan over the $(\beta,f_{WW})$ parameter space, we compute the combined efficiency (defined in equation~\ref{eff-def}) of the basic trigger level cuts on jets and leptons as 
well as the subsequent ATLAS cuts listed in tables~\ref{effOF} and \ref{effSF}, and it is well-fit by the following function
\begin{equation}
\epsilon_{WW^*+\geq 2{\rm -jets}} = \frac{50.98 \beta^4 + 121.76 \beta^3 f_{WW} + 22.85 \beta^2 f_{WW}^2 + 0.15 \beta f_{WW}^3 + 0.01 f_{WW}^4}
                          {1601.43 \beta^4 + 3796.63 \beta^3 f_{WW} + 666.79 \beta^2 f_{WW}^2 - 1.98 \beta f_{WW}^3 + 0.73 f_{WW}^4}.
\label{eff-ww}                          
\end{equation}
\begin{figure}[h!]
\begin{center}
\centerline{\epsfig{file=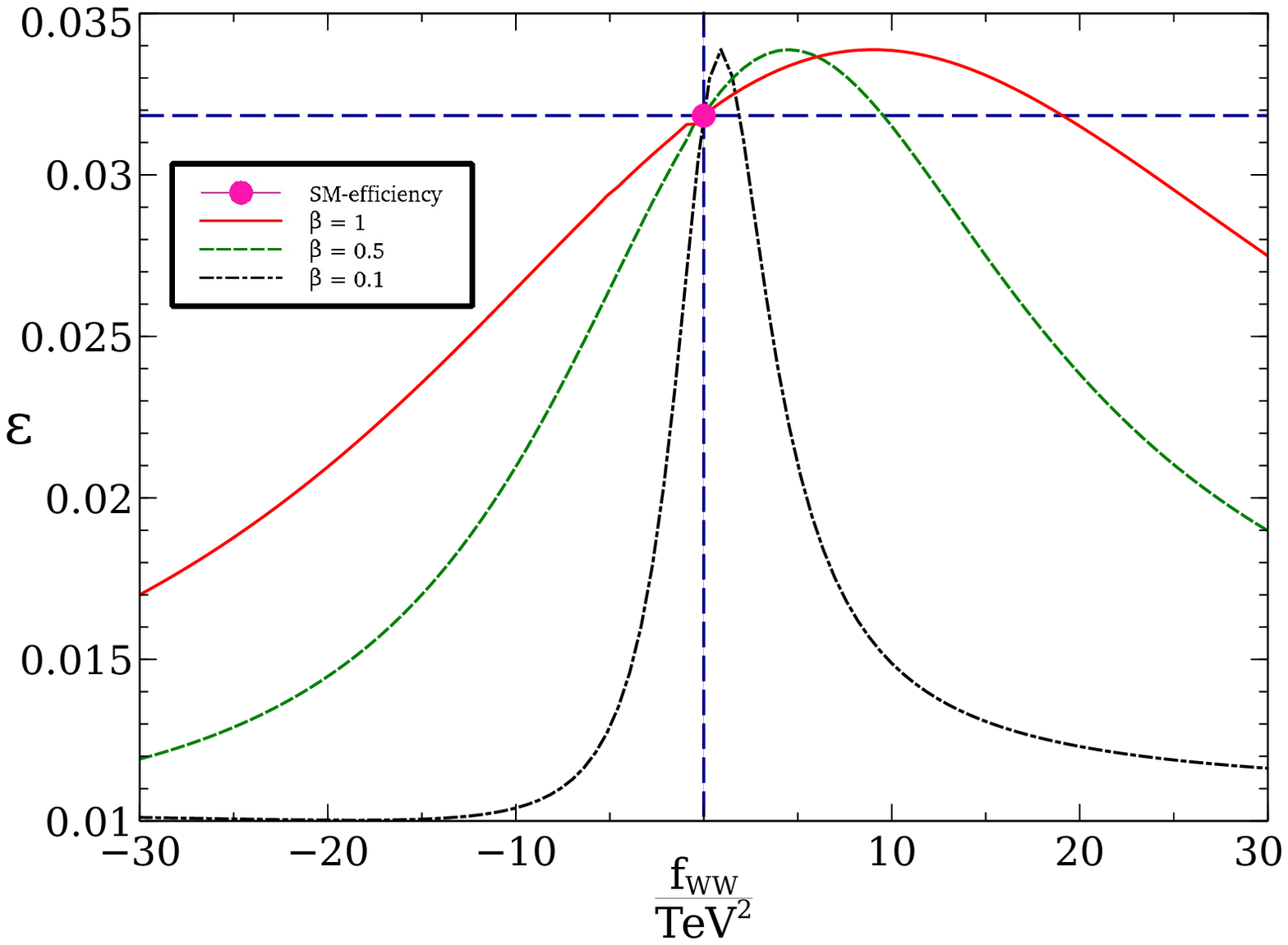
,width=12cm,height=10cm,angle=-0}}

\caption{\small \sl The combined efficiency of all ATLAS cuts ($\epsilon$) as a function of $f_{WW}$ for different values of $\beta$, in the $H\rightarrow WW^*\rightarrow \ell^+ \nu \ell^- \bar{\nu}$ channel ($\geq 2$-jets category) at 8 TeV LHC.}
\label{fig:eff-vs-f}
\end{center}
\end{figure}
In figure~\ref{fig:eff-vs-f} we show the variation of $\epsilon_{WW^*+\geq 2{\rm -jets}}$ as a function of $f_{WW}$ for different values of $\beta$. The red (solid), green (dashed) and black 
(dot-dashed) curves correspond to $\beta=1,0.5{~\rm and}~0.1$ respectively. For $f_{WW}=0$, we recover the SM efficiency ($\epsilon_{\rm SM}\simeq 0.032$) for all values of $\beta$. This fact 
confirms our expectation that only the introduction of new Lorentz structures changes the efficiencies, and a scaling  of the SM coupling alone by the factor $\beta$ does not. However, as we 
can see from this figure, although the overall features of the three curves are similar, for different values of $\beta$, the change in slopes are markedly different. Within the range of 
$f_{WW}$ shown in this figure, for $\beta=0.5$, the efficiency can reduce from its SM value by upto a factor of $2.5$ or more, while for $\beta=0.1$, it can drop by upto a factor of $3$.

\begin{figure}[htb!]
\begin{center}
\centerline{\epsfig{file=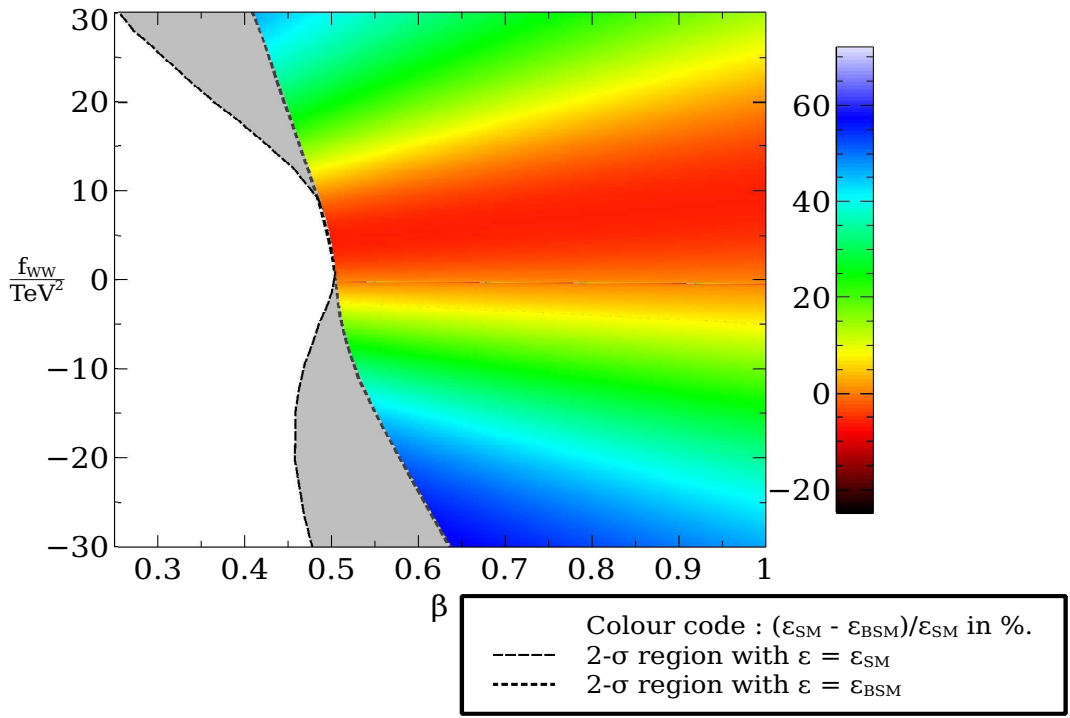,width=14cm,height=10cm,angle=-0}}

\caption{\small \sl Percentage modification of the combined efficiency of all cuts compared to the SM case, in the $H\rightarrow WW^*\rightarrow \ell^+ \nu \ell^- \bar{\nu}$ channel for the 
$\geq 2$-jets category at 8 TeV LHC. Only the region allowed at $95\%$ C.L. after imposing the ATLAS signal-strength constraint in this channel is shown. For comparison, we also show the 
allowed region at $95 \%$ C.L. (grey shaded region to the right of the dashed curve), with the assumption $\epsilon_{\rm BSM}=\epsilon_{\rm SM}$. }
\label{fig:eff-diff}
\end{center}
\end{figure}
Combining equations~\ref{eff-def} --\ref{eff-ww}, we can now evaluate the signal strength $\mu$ for the $H\rightarrow WW^*$ mode in the $\geq 2$-jets category, for any value of $\beta$ and 
$f_{WW}$. We emphasise that this calculation of the signal strength takes into account all the effects of the experimental cuts, and the resulting modification of their efficiencies compared 
to the SM case. This is one of our main results. In figure~\ref{fig:eff-diff} we show the percentage difference of $\epsilon_{\rm BSM}$ and $\epsilon_{\rm SM}$ in this channel in the 
$\beta-f_{WW}$ plane. The ranges of the parameters have been restricted to a region consistent at $95 \%$ C.L., with the signal strength measured in this channel by ATLAS 
($\hat{\mu}=1.66 \pm 0.79$)~\cite{ATLAS-WW}.  For comparison, we also show the allowed region at $95 \%$ C.L. (grey shaded region to the right of the dashed curve), with the assumption 
$\epsilon_{\rm BSM}=\epsilon_{\rm SM}$. It is clear that taking the efficiency modification into account significantly changes the parameter space allowed by the measurement in this channel. 
As we can also see from this figure, if this channel is considered on its own, the allowed region includes parameter points where the change in efficiency can be as large as $60 \%$. 
Therefore, in a completely rigorous global analysis of the data, this modification should be taken into account. 

\section{Constraints using LHC Higgs data}
\label{sec-5}
In the previous section, we have seen that the operator $\mathcal{O}_{WW}$ significantly modifies the final state kinematics in the $H+2$-jets channel (with $H\rightarrow WW^*$) and that a 
large region in the $\beta-f_{WW}$ parameter space is allowed by the current ATLAS measurement in this particular final state. However, the signal strengths in other bosonic channels are 
modified by $\mathcal{O}_{WW}$ as well. In this section, we therefore study the modifications to the inclusive $H \rightarrow W W^{*} ,H \rightarrow Z Z^{*} $ and 
$H \rightarrow \gamma \gamma $ channels in presence of non-zero $\beta$ and $f_{WW}$, and determine the most stringent possible constraints on these parameters. Before performing a global 
analysis with all the data taken together, we first analyze the constraints coming from each channel separately, in order to acquire a qualitative understanding. The signal strengths measured 
by the ATLAS and CMS collaborations, and a combination of the two experiments (assuming they are statistically independent) are shown in table~\ref{tab:alldata}. For the 
$H \rightarrow WW^{*} + 2{\rm -jets}$ channel, only the ATLAS result is available at present.
\begin{table}[htb!]
\centering
\begin{tabular}{|l|c|c|c|}
\hline
Channel & ATLAS & CMS & Combined\\
\hline

$H \rightarrow \gamma \gamma$ & $1.55^{+0.33}_{-0.28}$ & $0.77^{+0.27}_{-0.27}$ & $1.11^{+0.20}_{-0.20}$ \\
\hline
$H \rightarrow W W^{*}$ & $0.99^{+0.31}_{-0.28}$ & $0.68^{+0.20}_{-0.20}$ & $0.78^{+0.16}_{-0.16}$ \\
\hline
$H \rightarrow Z Z^{*}$ & $1.43^{+0.40}_{-0.35}$ & $0.92^{+0.28}_{-0.28}$ & $1.10^{+0.22}_{-0.22}$ \\
\hline
$H \rightarrow WW^{*} + 2{\rm -jets}$ & $1.66^{+0.79}_{-0.79}$ & NA & NA \\
\hline

\end{tabular}
\caption{\small \sl Signal strengths measured by the ATLAS and CMS collaborations, and a combination of the two experiments (assuming they are statistically independent) for the bosonic final 
states. For the $H \rightarrow WW^{*} + 2{\rm -jets}$ channel, only the ATLAS result is available at present.}
\label{tab:alldata}
\end{table}

\begin{figure}[htb!]
\begin{center}
\centerline{\epsfig{file=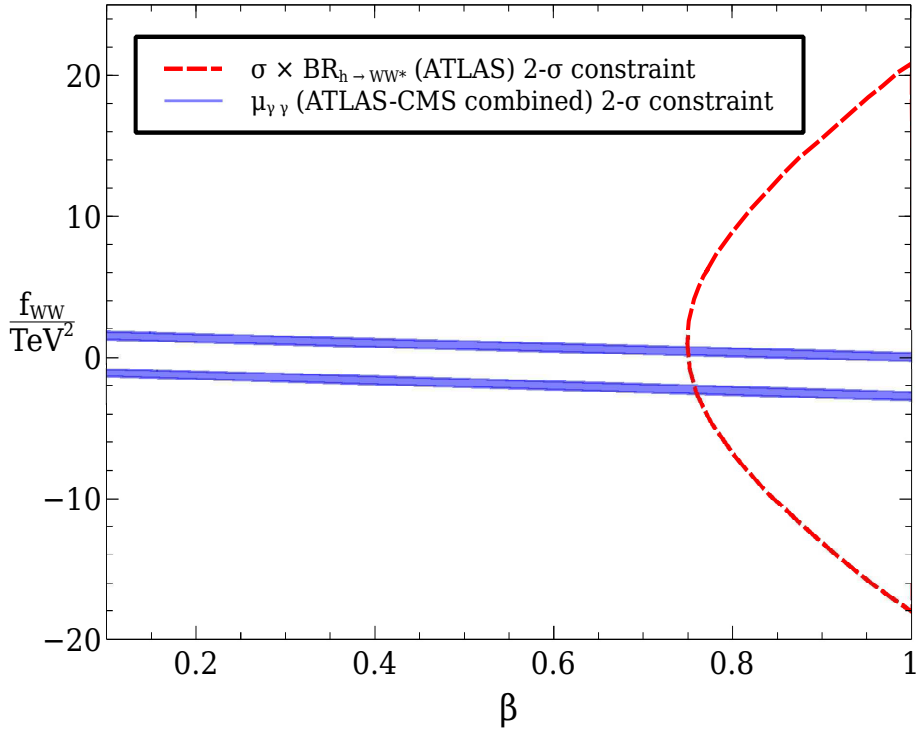
,width=12cm,height=10cm,angle=-0}}

\caption{\small \sl $2\sigma $ allowed regions in the $\beta-f_{WW}$ parameter space, after imposing the inclusive $\sigma(pp\rightarrow H)\times {\rm BR}(H\rightarrow WW^*)$ cross-section 
measurement by ATLAS (to the right of the red dashed curve), and  the combined ATLAS and CMS signal strength constraint  in the $\gamma \gamma$  channel (blue-shaded bands).}
\label{fig:gaga-ww-total}
\end{center}
\end{figure}

A measurement of the inclusive cross-section at 8 TeV LHC in the $WW^*$ channel has also been reported by ATLAS, after unfolding all detector effects, and it is found to be 
(for $m_H=125$ GeV)~\cite{ATLAS-WW}
\begin{equation}
\sigma(pp\rightarrow H)\times {\rm BR}(H\rightarrow WW^*)=6.0 \pm 1.6 {~\rm pb},
\end{equation}
which is slightly less than the expected SM cross-section ($4.8 \pm 0.7$ pb), but consistent with it within the uncertainties. We find that this measurement of the inclusive cross-section 
puts a severe constraint in the $\beta-f_{WW}$ parameter space, and the $2\sigma $ allowed region after imposing this requirement is shown in figure~\ref{fig:gaga-ww-total}. In the allowed 
region, to the right of the red (dashed) curve, $f_{WW}/\Lambda^2$ can be in the range $[-18:21]{~\rm TeV}^{-2}$, while $\beta$ is restricted to the range $[0.75:1.0]$. As we can see from equation~\ref{partial}, the relative magnitudes of the $\beta^2,\beta f_{WW}$ and $f_{WW}^2$ terms are similar for both $\Gamma_{H\rightarrow W W^{*}}$ and $\Gamma_{H\rightarrow Z Z^{*}}$. Therefore, their deviations from the SM will restrict $\beta$ and $f_{WW}$ in a similar range as well, especially since the signal strength measurements in both these channels have similar errors at the moment. Hence, we do not show the effect of the $ZZ^*$ channel separately, although it is included in our global fit to the data in bosonic channels.

As seen in equation~\ref{coupling-change}, $\mathcal{O}_{WW}$ also affects the $H\gamma \gamma$ coupling, and therefore, the inclusive signal-strength measured in this channel. 
Since gluon fusion is the dominant production mechanism for this mode, and we do not find significant deviations in the kinematics if the momentum-dependent couplings appear only in the decay 
vertices, an appreciable change in the cut-efficiency factor is not expected for this channel, and we do not include a modified $\epsilon_{BSM}$. In figure~\ref{fig:gaga-ww-total}, along with 
the inclusive total cross-section constraint in the $WW^*$ channel discussed above, we show the $2\sigma$ region, allowed by the combined signal strength measurement by ATLAS and CMS, 
$\hat{\mu}_{\gamma \gamma}=1.11 \pm 0.2$. Only the two blue-shaded regions are allowed by the current data, restricting the values of $f_{WW}$ to two narrow bands. For example, for $\beta=1$ 
the allowed values of $f_{WW}/(1{~\rm TeV}^2)$ are in the two sub-regions $[-3.05,-2.46]$ and $[-0.25,0.35]$. 
It is interesting to note that the intermediate region $-2.46 < f_{WW}/(1{~\rm TeV}^2)< -0.25$ is not allowed by the $2\sigma$ constraint. This is because, from equation~\ref{partial}, 
we can see that $\Gamma_{H\rightarrow \gamma \gamma}$ has a minimum at $f=-1.36$ for $\beta=1$. Therefore, in this intermediate region around the minimum, the signal strength becomes lower 
than the $2\sigma$ allowed lowest value. Similarly, for $\beta=0.1$, the allowed ranges for $f_{WW}/(1{~\rm TeV}^2)$ are $[-1.39,-0.85]$ and $[1.24,1.80]$, and the minimum of 
$\Gamma_{H\rightarrow \gamma \gamma}$ is at $f=0.21$. Since the operator $\mathcal{O}_{BB}$ modifies the $H \gamma \gamma$ coupling in exactly the same form as $\mathcal{O}_{WW}$ 
(see equation~\ref{coupling-change}), these constraints on $f_{WW}$ from the $\gamma \gamma$ data also apply to $f_{BB}$. The modified cut efficiencies in the $VH$ channel in presence of 
$\mathcal{O}_{BB}$ are studied in section~\ref{sec-6}.

\begin{figure}[htb!]
\begin{center}
\centerline{\epsfig{file=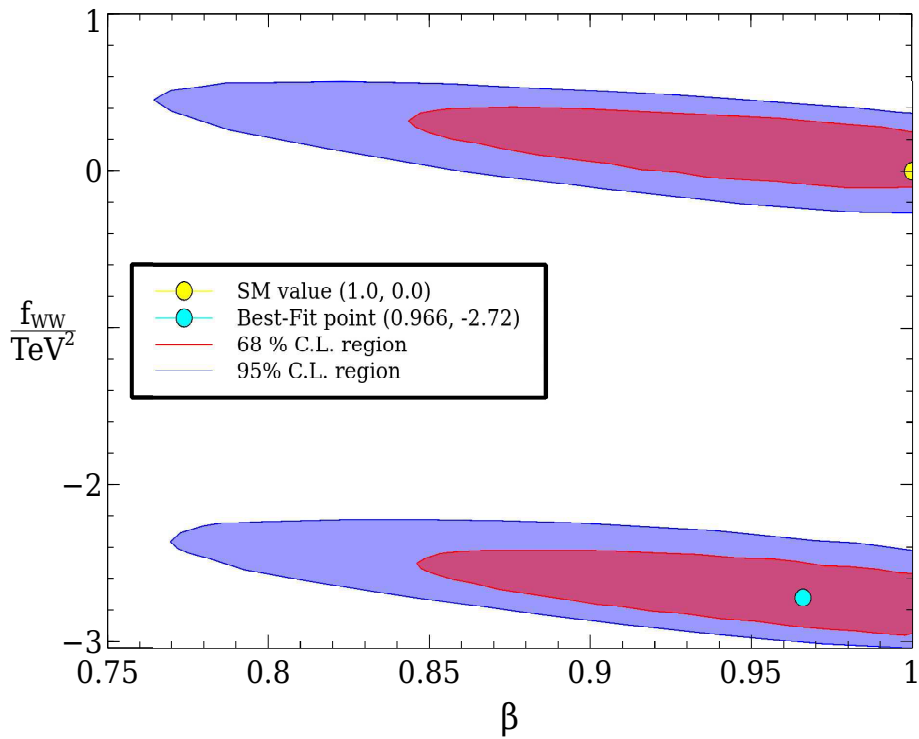,width=12cm,height=10cm,angle=-0}}

\caption{\small \sl  $68\%$ and $95\%$ C.L.  allowed regions in the $\beta-f_{WW}$ parameter space, after performing a global fit using the data in all bosonic channels given in 
table~\ref{tab:alldata}. The best-fit and SM points are also shown.}
\label{fig:global-fit}
\end{center}
\end{figure}
As we can see from equation~\ref{partial}, for the $\gamma \gamma$ partial width, the contribution from $\mathcal{O}_{WW}$ is comparable in magnitude to the loop-induced W-boson contribution, 
and therefore, values of $\beta$ as small as $0.1$ with $|f_{WW}/(1{~\rm TeV}^2)|<2$ are allowed by this constraint. This is not true for the $WW^*$ inclusive cross-section constraint, where, 
the SM-like term contributes 2 orders of magnitude higher (see $\Gamma_{H\rightarrow W W^*}$ in equation~\ref{partial}), thereby restricting $\beta$ to 0.75. Therefore, by comparing the 
$\gamma \gamma$ and $WW^*$ inclusive constraints taken separately, we can learn that while large values of $f_{WW}$ is disallowed by the former, small values of $\beta$ are ruled out by the 
latter. A combination of all the constraints brings us to the global analysis using the data in the bosonic channels (see table~\ref{tab:alldata}), the result of which is presented in 
figure~\ref{fig:global-fit}. The constraints on each of the parameters coming from the  global fit is now easily understood in terms of the arguments given above. 
The best fit point corresponds to $\beta=0.97$ and $f_{WW}/(1{~\rm TeV}^2)=-2.72$, which are very close to the SM point. However, there is still a small room for new physics effects described 
by $\mathcal{O}_{WW}$ and $\beta$, as can be seen from the allowed regions at the $2\sigma$ level. 
\begin{figure}[htb!]
\begin{center}
\centerline{\epsfig{file=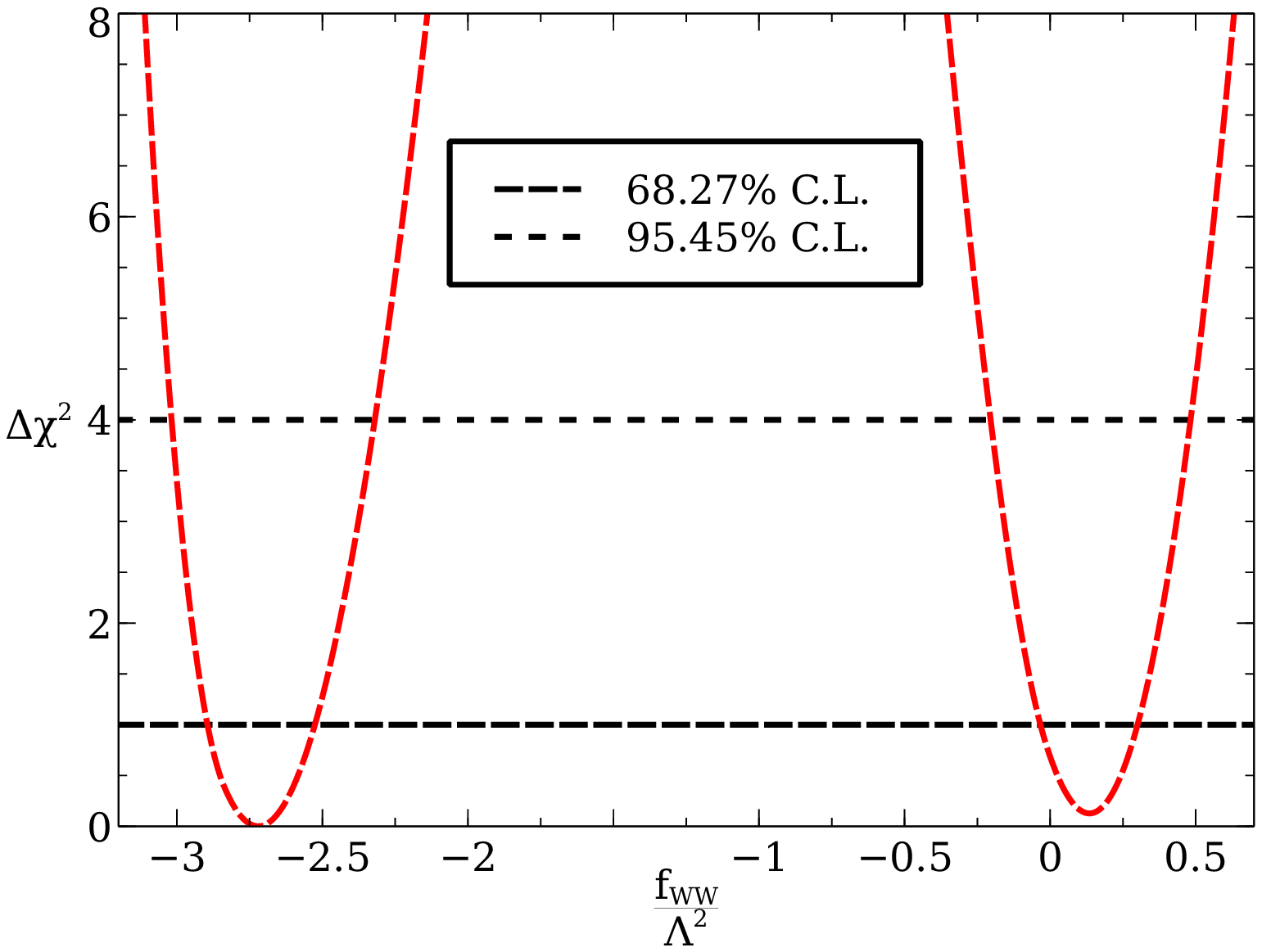,width=8cm,height=7cm,angle=-0}
\epsfig{file=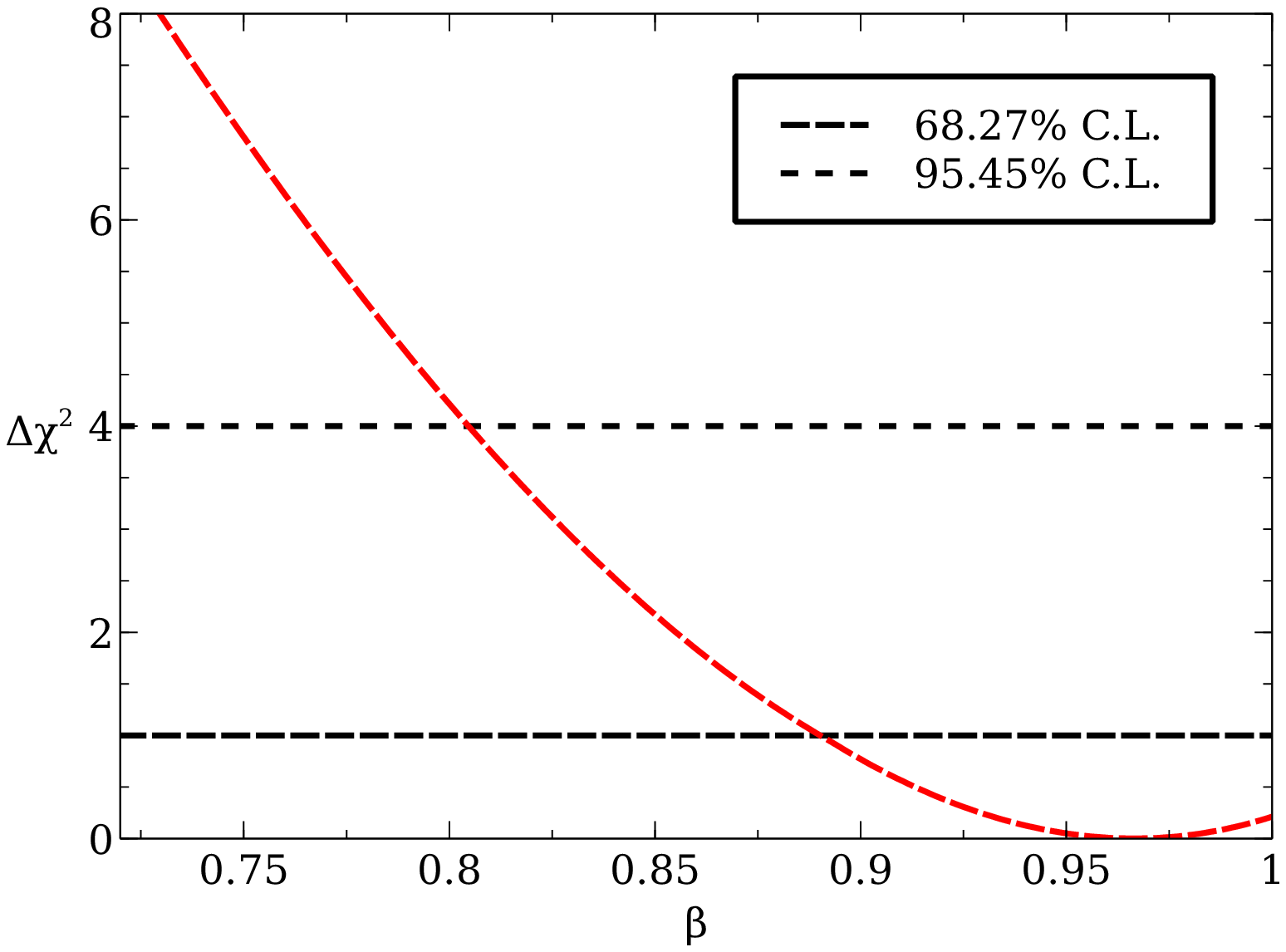,width=8cm,height=7cm,angle=-0}}

\caption{\small \sl $\Delta \chi^2$ distributions as a function of $f_{WW}$ (left) and $\beta$ (right), after marginalizing over $\beta$ and $f_{WW}$ respectively. The allowed ranges at 
$68\%$ and $95\%$ C.L. are also shown by the horizontal dashed lines.}
\label{margin}
\end{center}
\end{figure}
We also show in figure~\ref{margin} the $\Delta \chi^2$ distributions as a function of $f_{WW}$ and $\beta$, after marginalizing over $\beta$ and $f_{WW}$ respectively. From this figure, we 
obtain the allowed range for $\beta$ as
\begin{equation}
      0.8\leq  \beta \leq 1.0  ~~~~~~~95\% {~\rm C.L., marginalized ~over~} f_{WW}.
\end{equation}
Similarly, the allowed range for $f_{WW}$ is found to be
\begin{equation}
\frac{f_{WW}}{{~\rm TeV}^{-2}} \in [-3.02,-2.31]\cup[-0.20,0.48] ~~~~~~~95\% {~\rm C.L., marginalized ~over~} \beta.
\end{equation}
As far as the dimension-six operator $\mathcal{O}_{WW}$ is concerned, this bound tells us that for a suppression scale $\Lambda=1$ TeV, the co-efficient $f_{WW}$ cannot be smaller than 
$\sim -3$, or, in other words, if $f_{WW}=\mathcal{O}(1)$, $\Lambda \gtrsim 600$ GeV. This result is consistent with the present LHC direct search bounds on the mass of  new coloured or 
uncoloured particles charged under the electroweak gauge group. 

Two points are worth mentioning here. Firstly, it is true that the apparently allowed range of $f_{WW}$ after fitting the data suggest rather modest change in cut efficiencies due to the 
presence of the additional operator. Still we consider the general demonstration of altered efficiencies over a range of the parameter space, quite substantially different in some cases, to be useful. Such altered efficiencies may plague our results, if, for example, overconstraining of $f_{WW}$ has taken place because of event migration. Moreover, in our global fit, we could only include the modified cut efficiencies in the $WW^*+\geq 2-$jets channel, and not in the other important bosonic channels like $\gamma \gamma$, $ZZ^*$ and $WW^*$ (inclusive), as the detailed information on the cut-flows for the latter channels are not yet presented by the experimental collaborations. Thus the $95\%$ C.L. allowed regions obtained by a global fit with our current set-up is very similar to the region obtained using SM efficiencies. Once the detailed information of experimental cut-flows in all channels is available, our method can be extended to perform a fully rigorous global analysis. Secondly, even within the `apparently 
allowed' range, it is worthwhile to look for modification in kinematic distributions due to the additional operators. For example, variables used in Boosted Decision Trees (BDT) can still bear the stamp of the higher-dimensional operators, as we shall see in section~\ref{sec-7}.

\section{Associated production and higher dimensional operators}
\label{sec-6}
In section~\ref{sec-4}, we studied the change in efficiencies of the cuts used by ATLAS in the $H \rightarrow WW^{*}$ channel ($\geq2$-jet category), in presence of non-zero $\beta$ and 
$f_{WW}$. These cuts were tailored to primarily select $VBF$ events. In this section, we study the modification of cut-efficiencies in the associated production (VH) channel, in presence 
of $\mathcal{O}_{WW}$ as well as $\mathcal{O}_{BB}$, taking their effects one at a time. As the operator $\mathcal{O}_{BB}$ only modifies the $HZZ$ and $H\gamma\gamma$ vertices, but not 
the $HWW$ vertex (see equations~\ref{Interactions},~\ref{coupling-change}), its effect in the $VBF$ channel is not significant\footnote{In the SM, the WW-fusion diagram contributes roughly 
$3$ times to the VBF Higgs production cross-section, compared to the ZZ-fusion diagram, while the interference term is negligible ($\sim 1\%$)~\cite{Djouadi2}.}. 
We do not show the modification of efficiencies for the $H\rightarrow ZZ^{*}$ channel in the VH category, mainly because the cuts given in the corresponding experimental papers are not so 
transparently provided when compared to the $WW^{*}$ and $\gamma \gamma$ final states.

Thus we focus here on the $VH$ production of Higgs, where the Higgs decays to two photons and the vector boson ($W$ or $Z$) decays hadronically. We closely follow the cuts used by ATLAS 
(see table~\ref{tab:gaga}) and study modifications to their efficiencies. The photon isolation criteria have been required to be 
$\Delta R_{\gamma \gamma} \geq 0.3$, $\Delta R_{\gamma j} \geq 0.4$ and $\Delta R_{\gamma l} \geq 0.4$, where $j$ and $l$ denote jets and leptons ($e$, $\mu$), 
and $\Delta R_{ij} = \sqrt{(\eta_{i} - \eta_{j})^2 + (\phi_{i} - \phi_{j})^2}$, $\eta$ and $\phi$ being the pseudorapidity and azimuthal angle respectively. In addition, a photon 
is considered isolated only if the total transverse energy around it in a cone of size $\Delta R = 0.4$ is less than $6$ GeV. For further details on the cuts, we refer the reader 
to references~\cite{atlas3,atlas-gaga}.
\begin{savenotes}
\begin{table}[htb!]
\centering
\begin{tabular}{|l|}
\hline
$100$ GeV $< m_{\gamma \gamma} <$ $160$ GeV \\
\hline
$60$ GeV $< m_{jj} <$ $110$ GeV \\
\hline
$|\Delta y_{jj}| < 3.5$ \\
\hline
$|\Delta \eta_{\gamma \gamma, j j}| < 1$ \\
\hline
$p_{T_{t}} > 70$ GeV \footnote{$p_{T_{t}}$ is the diphoton transverse momentum orthogonal to the diphoton thrust axis in the transverse plane, as defined later in section~\ref{sec-7}.} \\
\hline
\end{tabular}
\caption{\small \sl Cuts used for the $VH$ channel with $H \rightarrow \gamma \gamma$ and $V \rightarrow jj$ in the low mass two-jet category (ATLAS). See references~\cite{atlas3,atlas-gaga} 
for details.}
\label{tab:gaga}
\end{table}
\end{savenotes}

As in equation~\ref{eff-def}, we define the efficiency for this channel as a function of $\beta$ and $f$ ($f_{WW}$ or $f_{BB}$) as 
\begin{align}
\epsilon_{\gamma \gamma + 2-{\rm jets} (VH)} (\beta,f_{WW/BB}) = \frac{\left[\sigma(pp \rightarrow H)_{\rm VH, \rm V \rightarrow j j } \times {\rm BR} (H \rightarrow \gamma \gamma)  \right]_{\rm After ~Cuts}}{\left[\sigma(pp \rightarrow H)_{\rm VH, \rm V \rightarrow j j } \times {\rm BR} (H \rightarrow \gamma \gamma)  \right]_{\rm Before ~Cuts}}.
\label{eff-def2}
\end{align}
By performing a scan over the ($\beta$,$f_{WW/BB}$) parameter space, we obtain the combined efficiencies (for $\mathcal{O}_{BB}$ and $\mathcal{O}_{WW}$) of the isolation 
cuts and the ATLAS cuts in table~\ref{tab:gaga} and they are well fit by the following functions :
\small
\begin{multline}
\epsilon_{\gamma \gamma + 2-{\rm jets} (VH)} (\beta, f_{BB}) = \\ \dfrac{(3.75 \beta^{2} +2.66 \beta f_{BB} + 0.47 f_{BB}^{2})\times(0.15 - 1.34 \beta + 0.01 f_{BB} - 1.22 \beta^{2} - 0.05 \beta f_{BB} - 1.3\times10^{-11} f_{BB}^{2})}{(5.20 \beta^{2} + 3.68 \beta f_{BB} + 0.65 f_{BB}^{2})\times(2.05 - 18.76 \beta + 0.06 f_{BB} - 17.57 \beta^{2} - 0.65 \beta f_{BB} - 1.4\times10^{-10} f_{BB}^{2})},
\label{eff-bb-gaga}    
\end{multline}
\normalsize
\small
\begin{multline}
\epsilon_{\gamma \gamma + 2-{\rm jets} (VH)} (\beta, f_{WW}) = \\ \dfrac{(15.46\beta^{2}-1.33\beta f_{WW}+0.05 f_{WW}^{2})\times(0.03-0.35\beta+0.05 f_{WW}-8.88\beta^{2}+61.25 \beta f_{WW}-15.31 f_{WW}^{2})}{(0.64 \beta^{2}-4.12\beta f_{WW}+1.03 f_{WW}^{2})\times(-1.33+11.22 \beta+0.22 f_{WW}-4346.94 \beta^{2}+392.44 \beta f_{WW}-14.86 f_{WW}^2)}.
\label{eff-ww-gaga}  
\end{multline}
\normalsize
In equations~\ref{eff-bb-gaga} and~\ref{eff-ww-gaga}, the denominator and the numerator represent  $\sigma^{VH}_{prod} \times B.R. (H \rightarrow \gamma \gamma)$,  
before and after all the cuts respectively, and some common numerical  factors between the two, like the total Higgs decay width have been cancelled out.
\begin{figure}[h!]
\begin{center}
\centerline{\epsfig{file=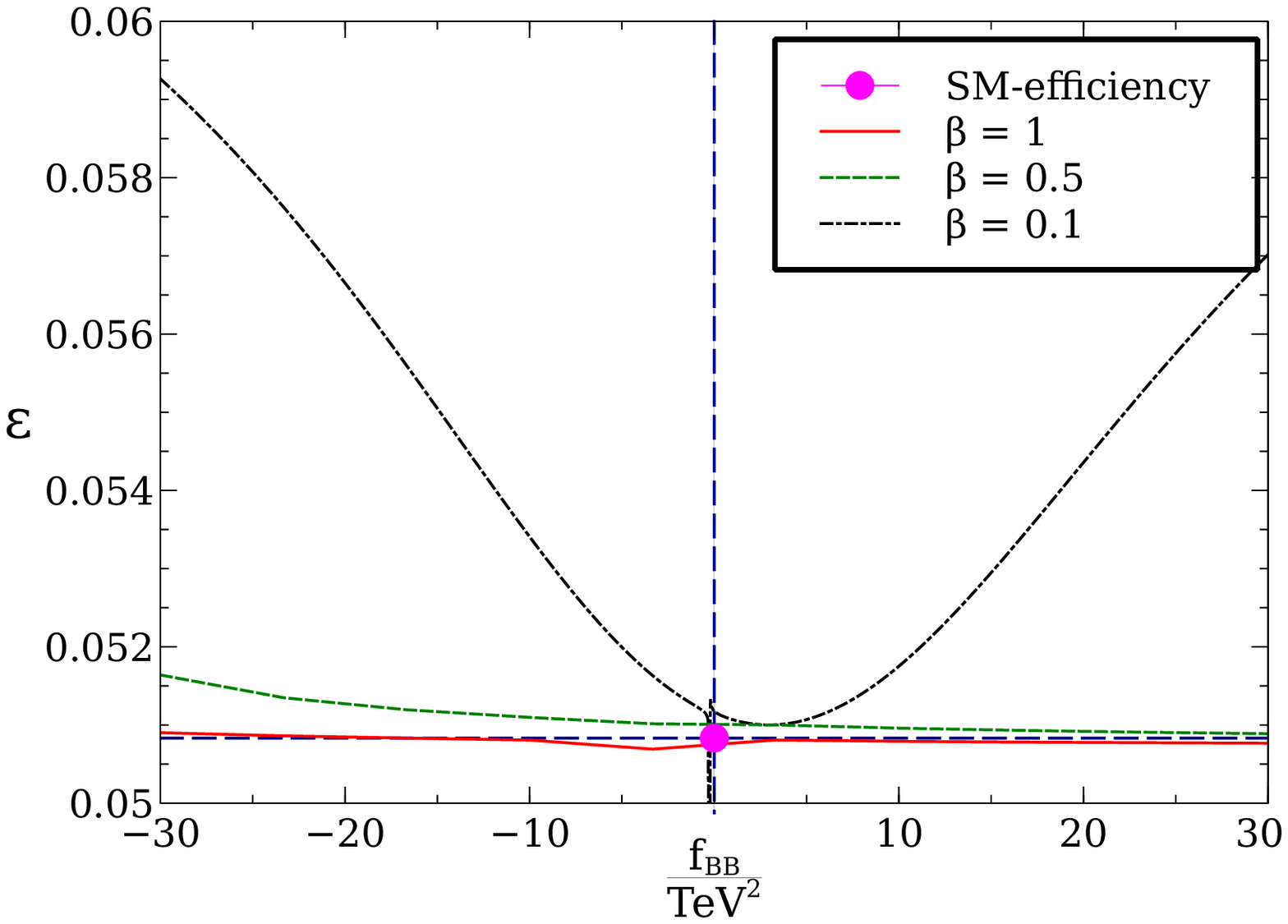,width=12cm,height=9.5cm,angle=-0}}
\caption{\small \sl The combined efficiency of all ATLAS cuts ($\epsilon$) as a function of $f_{BB}$ for different values of $\beta$, in the $H\rightarrow \gamma \gamma$ channel (VH category) at 8 TeV LHC.}
\label{fig:eff-vs-f-bb}
\end{center}
\end{figure}

\begin{figure}[h!]
\begin{center}
\centerline{\epsfig{file=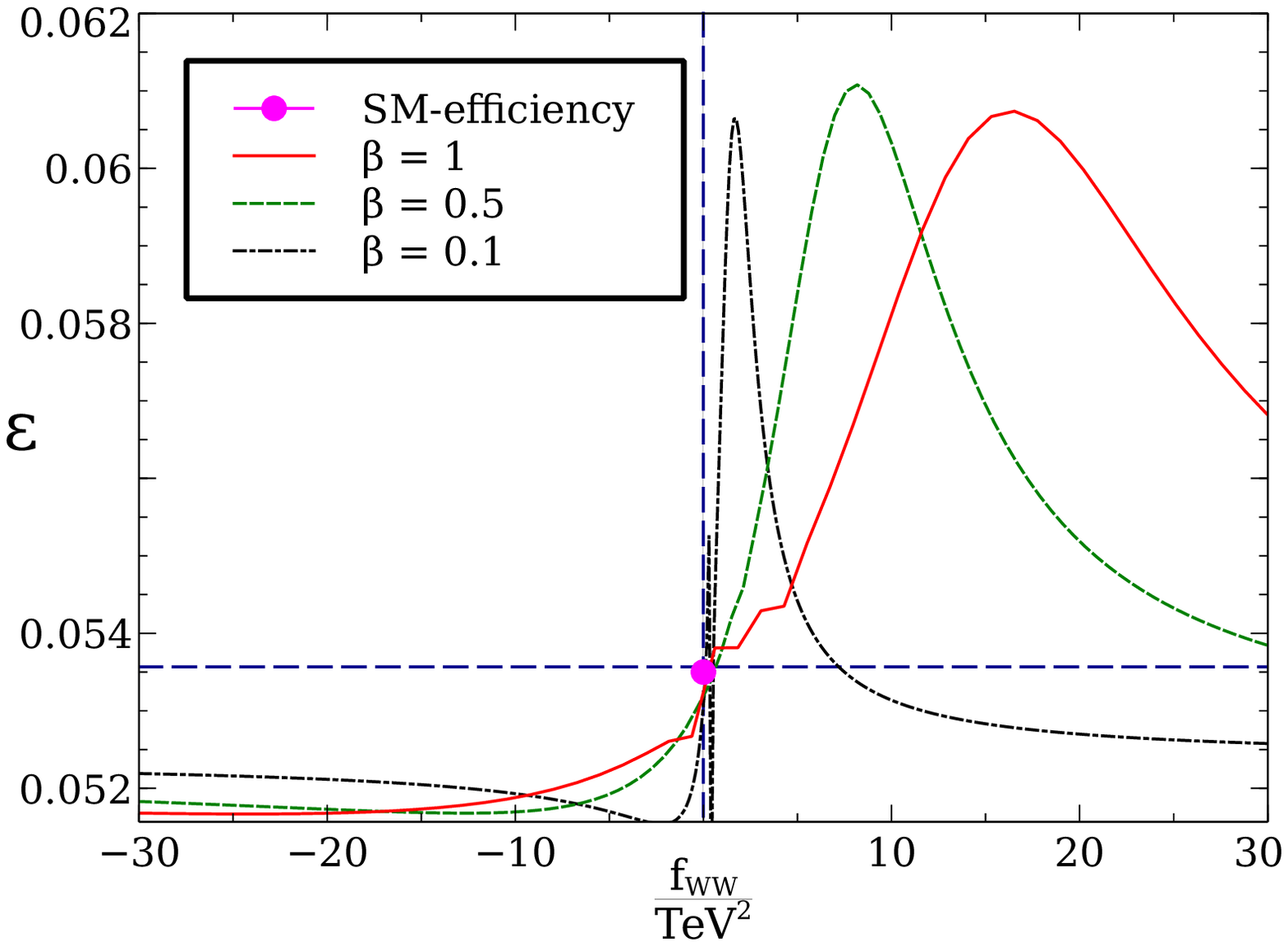,width=12cm,height=9.5cm,angle=-0}}

\caption{\small \sl The combined efficiency of all ATLAS cuts ($\epsilon$) as a function of $f_{WW}$ for different values of $\beta$, in the $H\rightarrow \gamma \gamma$ channel (VH category) at 8 TeV LHC.}
\label{fig:eff-vs-f-ww}
\end{center}
\end{figure}

In figures~\ref{fig:eff-vs-f-bb} and~\ref{fig:eff-vs-f-ww}, we show the variation of $\epsilon_{\gamma \gamma + 2-{\rm jets} (VH)}$ as a function of $f_{BB}$ and $f_{WW}$ respectively for 
different values of $\beta$. The red (solid), green (dashed) and black (dot-dashed) curves correspond to $\beta=1,0.5{~\rm and}~0.1$ respectively. For $f_{BB}=0$ or $f_{WW}=0$ we recover 
the SM efficiency ($\epsilon_{\rm SM}\sim 0.053$) in both the cases for all values of $\beta$. Within the range of $f_{BB}$ shown in figure~\ref{fig:eff-vs-f-bb}, the efficiency for 
$\beta = 0.5$ can change from its SM value by upto $8.7\%$, while for $\beta=0.1$, it can increase by upto $19.7\%$. On the other hand, for the range of $f_{WW}$ shown in 
figure~\ref{fig:eff-vs-f-ww}, for $\beta=0.5$ ($\beta=0.1$), the efficiency can increase from its SM value by upto $14.7\%$ $(13.9\%)$. Thus our overall conclusion is that the modification of 
cut efficiencies in the VH production mode is in general less pronounced than in the VBF Higgs production with $H \rightarrow WW^{*}$.

\section{Modification to kinematic distributions : examples}
\label{sec-7}
Since the modifications of cut efficiencies discussed so far originate from changes in kinematic distributions, in this section, we explore some of these distributions in the presence of 
higher dimensional operators. Study of differential distributions is the next step in experimental analysis of the Higgs sector, and preliminary results with the current data have already 
been presented in Ref~\cite{atlas4}. As an example we choose the diphoton channel in the VBF category~\cite{atlas3,atlas-gaga}, and consider the operator $\mathcal{O}_{WW}$ for illustration. 
All the distributions are shown after applying the standard trigger and isolation cuts for jets and photons. The kinematic variables considered are : 
\begin{itemize}
 \item[1.] $\sqrt{\vec{p}_{Tj_{1}} \cdot \vec{p}_{Tj_{2}}}$, where $j_{1}$ and $j_{2}$ are the two tagged jets ordered in terms of their transverse momenta.
 \item[2.] $|\Delta \eta_{j_{1}j_{2}}| = |\eta_{j_{1}} - \eta_{j_{2}}|$.
 \item[3.] The invariant mass of the two tagged jets, $m_{j_{1}j_{2}}$. For this as well as the distributions listed below, the cut $|\Delta \eta_{j_{1}j_{2}}| > 2.8$ 
           (see Ref.~\cite{atlas-gaga2}) is imposed.
 \item[4.] $p_{T_{t}}=|\vec{p}_{T}^{\gamma \gamma} \times \hat{t}|$, where 
       $\hat{t} = \frac{\vec{p}_{T}^{\gamma 1} - \vec{p}_{T}^{\gamma 2}}{|\vec{p}_{T}^{\gamma 1} - \vec{p}_{T}^{\gamma 2}|}$ is the transverse 
       thrust, $\vec{p}_{T}^{\gamma 1}$, $\vec{p}_{T}^{\gamma 2}$ are the transverse momenta of the two isolated photons and 
       $\vec{p}_{T}^{\gamma \gamma}=\vec{p}_{T}^{\gamma 1} + \vec{p}_{T}^{\gamma 2}$ is the transverse momentum of the diphoton system~\cite{atlas3,atlas-gaga}. This and the subsequent 
       distributions are subjected to the cuts $m_{j_{1}j_{2}} > 400$ GeV and $\Delta \phi_{\gamma \gamma, j_{1}j_{2}} > 2.6$, where $\Delta \phi_{\gamma \gamma, j_{1}j_{2}}$ is the azimuthal angle separation 
       between the diphoton system and the system of the two tagged jets. The criterion of no hadronic activity in the rapidity gap between the two tagged jets is also 
       imposed~\cite{atlas-gaga2}.
 \item[5.] $\eta^{*} = \eta_{\gamma \gamma} - \frac{\eta_{j1} + \eta_{j2}}{2}$, where $\eta_{\gamma \gamma}$ is the pseudorapidity of the diphoton system.
 \item[6.] $\Delta R^{\gamma j}_{\rm{min}}$ is the minimal $\Delta R$ between a photon and a tagged jet. 
\begin{figure}[htb!]
\centering
\centerline{\includegraphics[width=8.5cm,height=8.5cm,keepaspectratio]{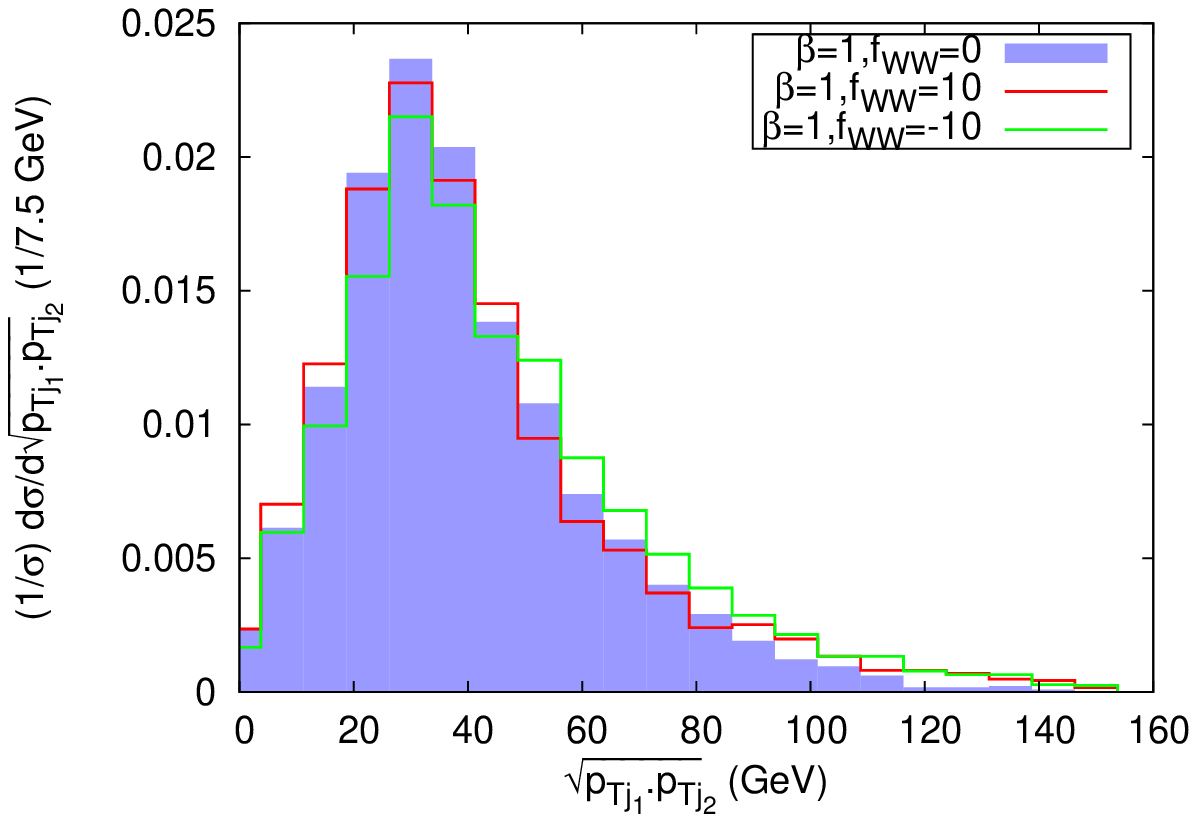}\
\includegraphics[width=8.5cm,height=8.5cm,keepaspectratio]{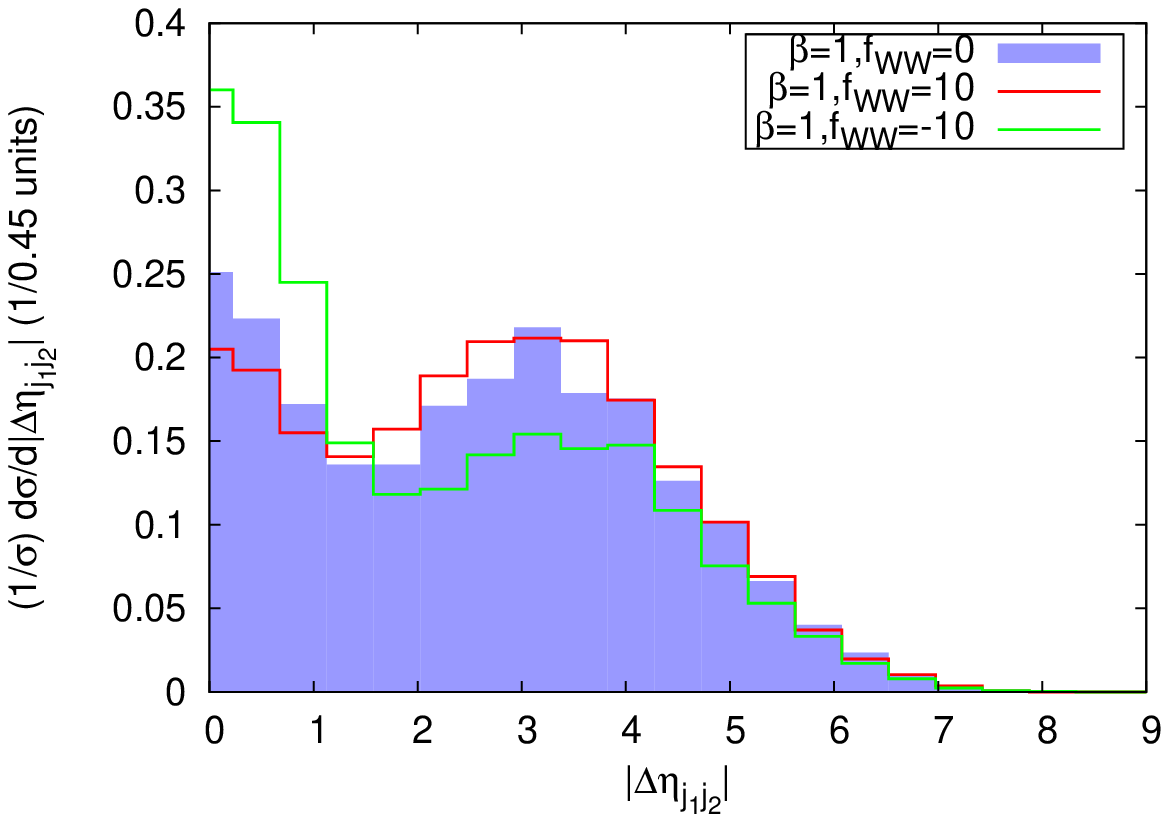}}
\centerline{\includegraphics[width=8.5cm,height=8.5cm,keepaspectratio]{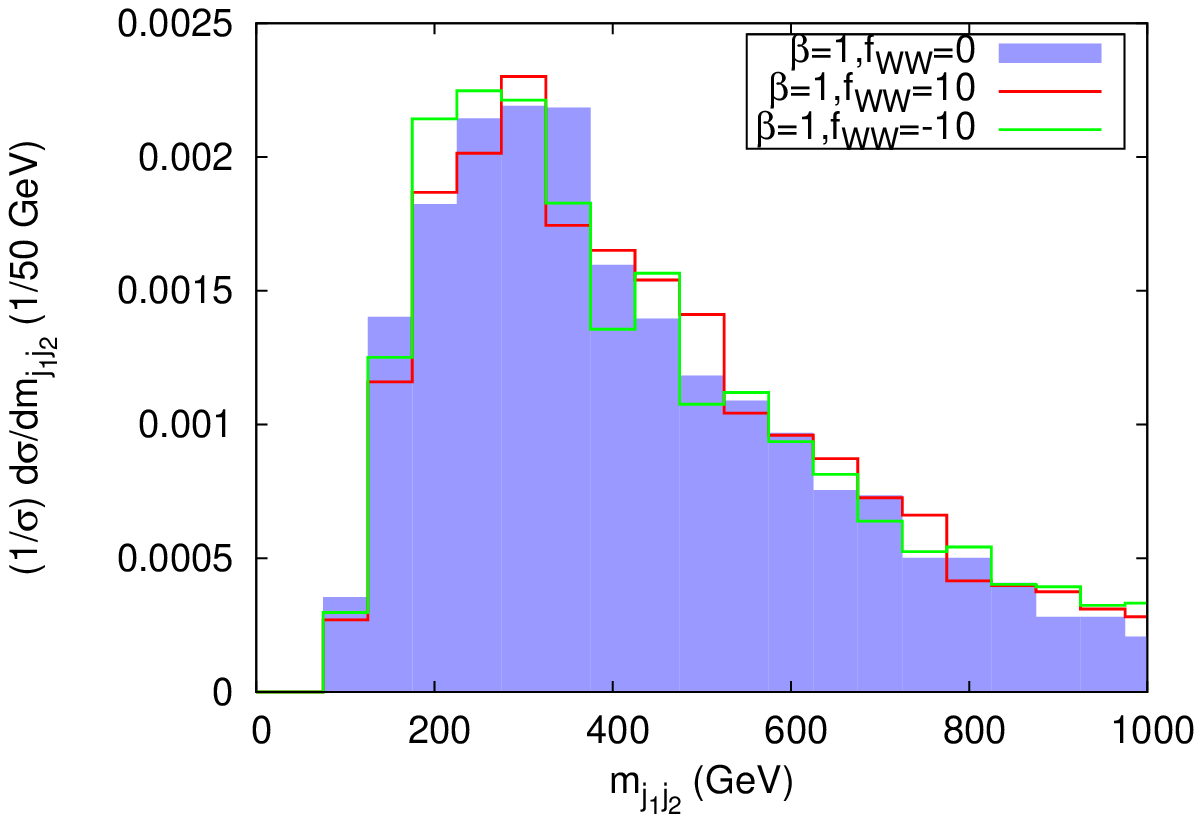}\
\includegraphics[width=8.5cm,height=8.5cm,keepaspectratio]{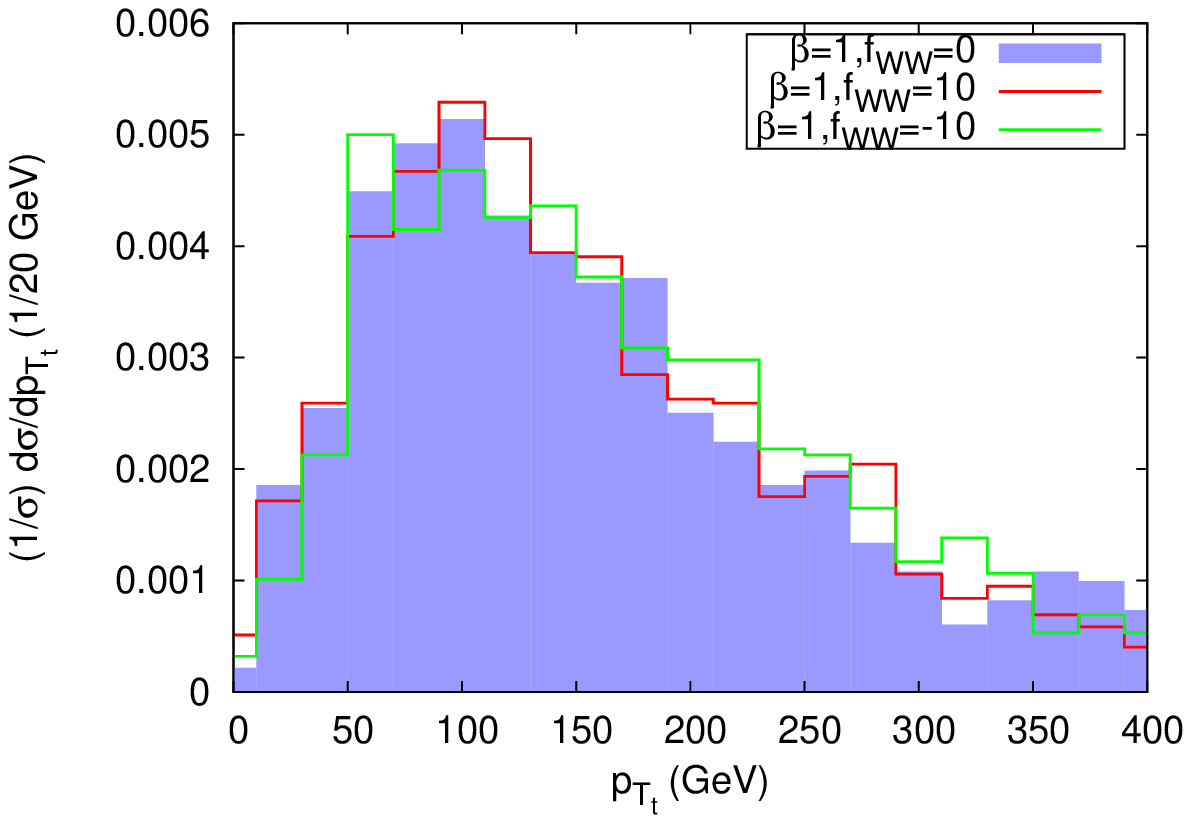}}
\centerline{\includegraphics[width=8.5cm,height=8.5cm,keepaspectratio]{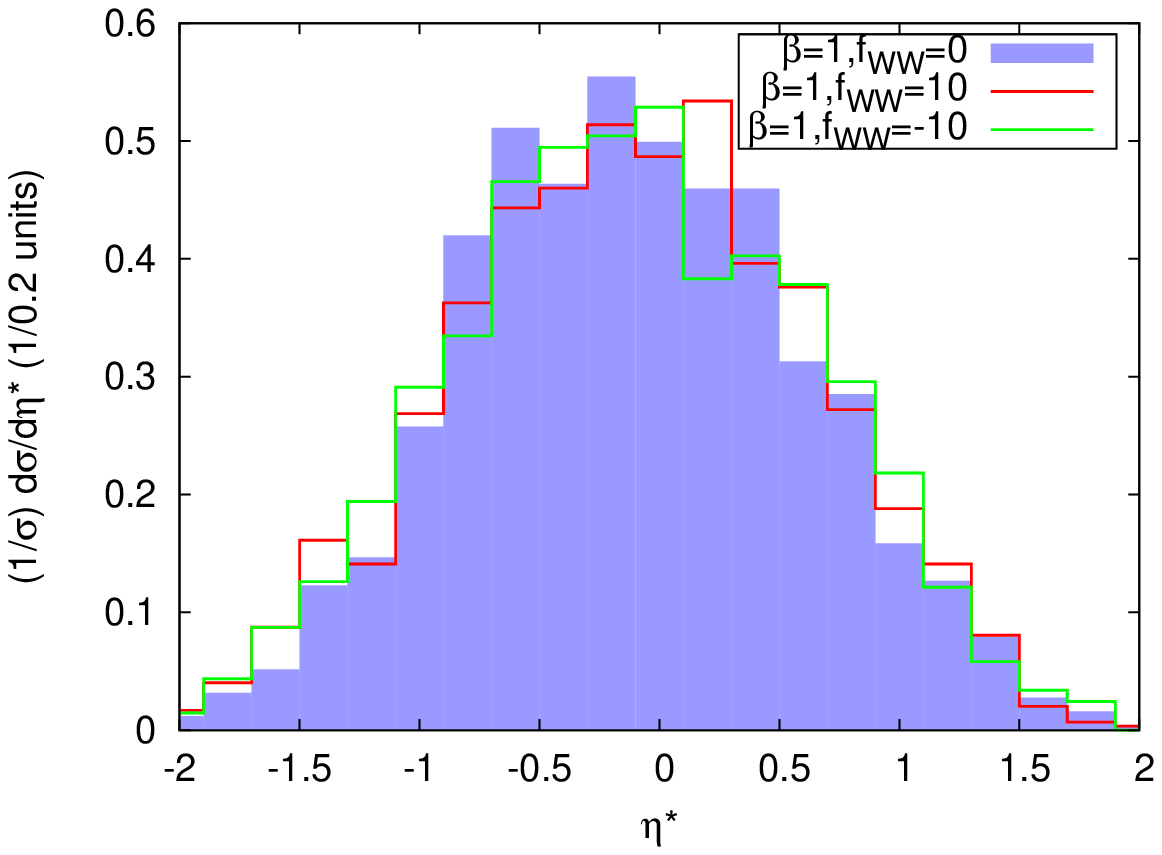}\
\includegraphics[width=8.5cm,height=8.5cm,keepaspectratio]{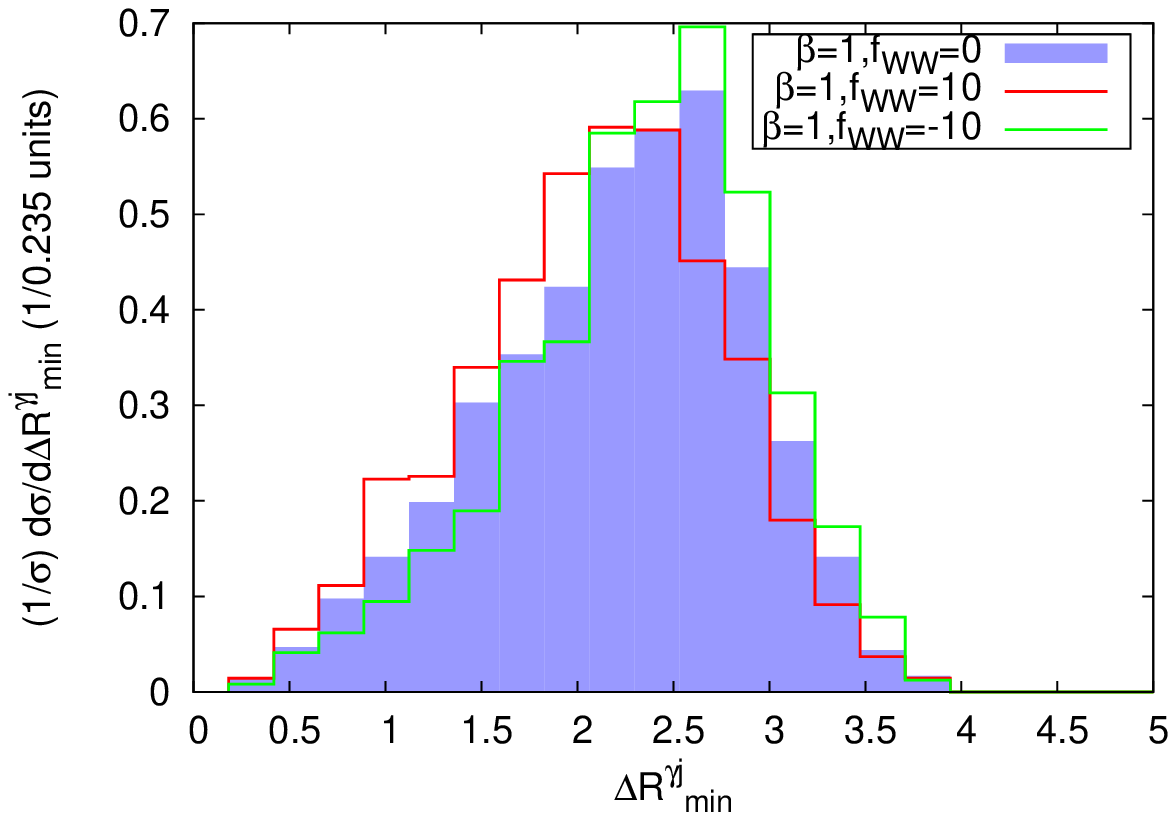}}
\caption{\small \sl Normalised distributions in various kinematic variables for 8 TeV LHC, top row : $\sqrt{\vec{p}_{Tj_{1}} \cdot \vec{p}_{Tj_{2}}}$ (left) and 
                    $|\Delta \eta_{j_{1} j_{2}}|$ (right); middle row : $m_{j_{1} j_{2}}$  (left) and $p_{T_{t}}$ (right); bottom row : $\eta^{*}$ (left) and 
                    $\Delta R_{\rm min}^{\gamma j}$ (right), for the parameter points $\{\beta = 1,f_{WW} = 0\}$ (shaded 
                    blue region), $\{\beta = 1,f_{WW} = 10\}$ (solid red line) and $\{\beta = 1,f_{WW} = -10\}$ (solid green line). The cut-off scale chosen is $\Lambda=1$ TeV.}
\label{fig:dist-b1-f0pm10}
\end{figure}
\begin{figure}[htb!]
\centering
\centerline{\includegraphics[width=8.6cm,height=8.6cm,keepaspectratio]{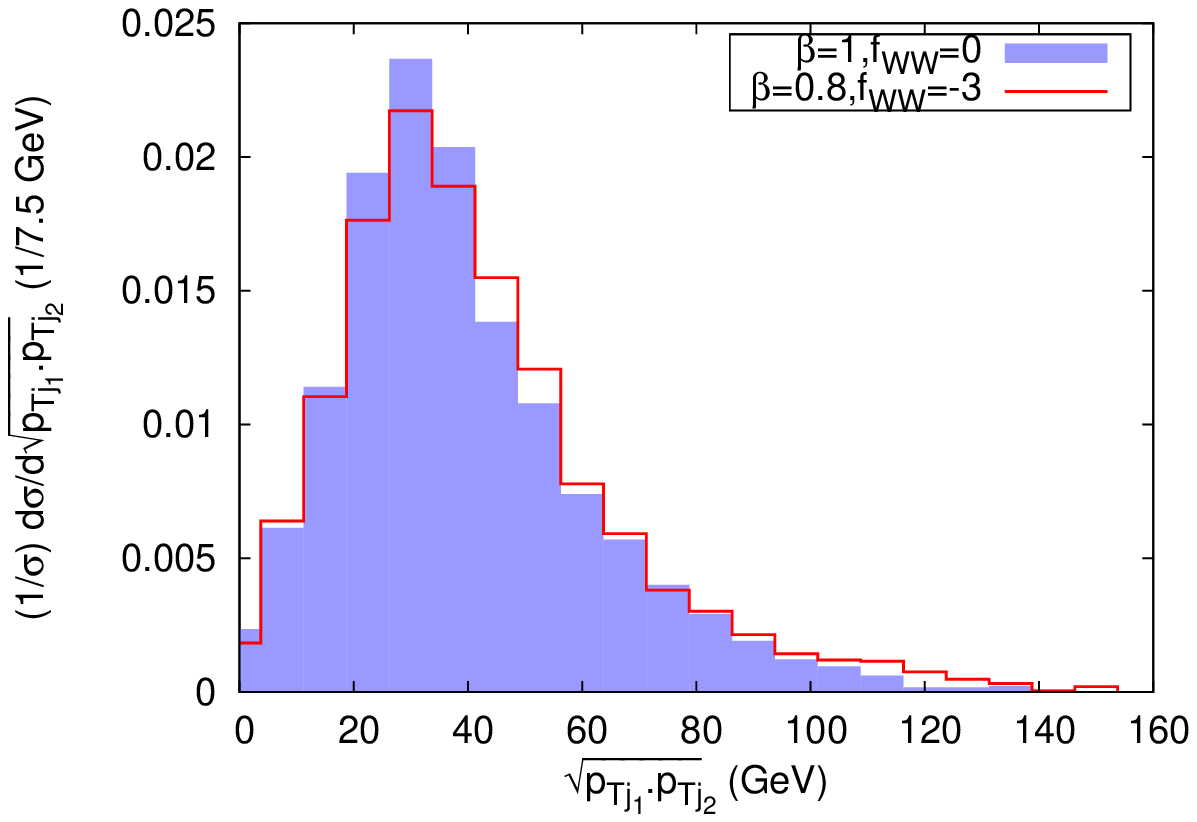}\
\includegraphics[width=8.6cm,height=8.6cm,keepaspectratio]{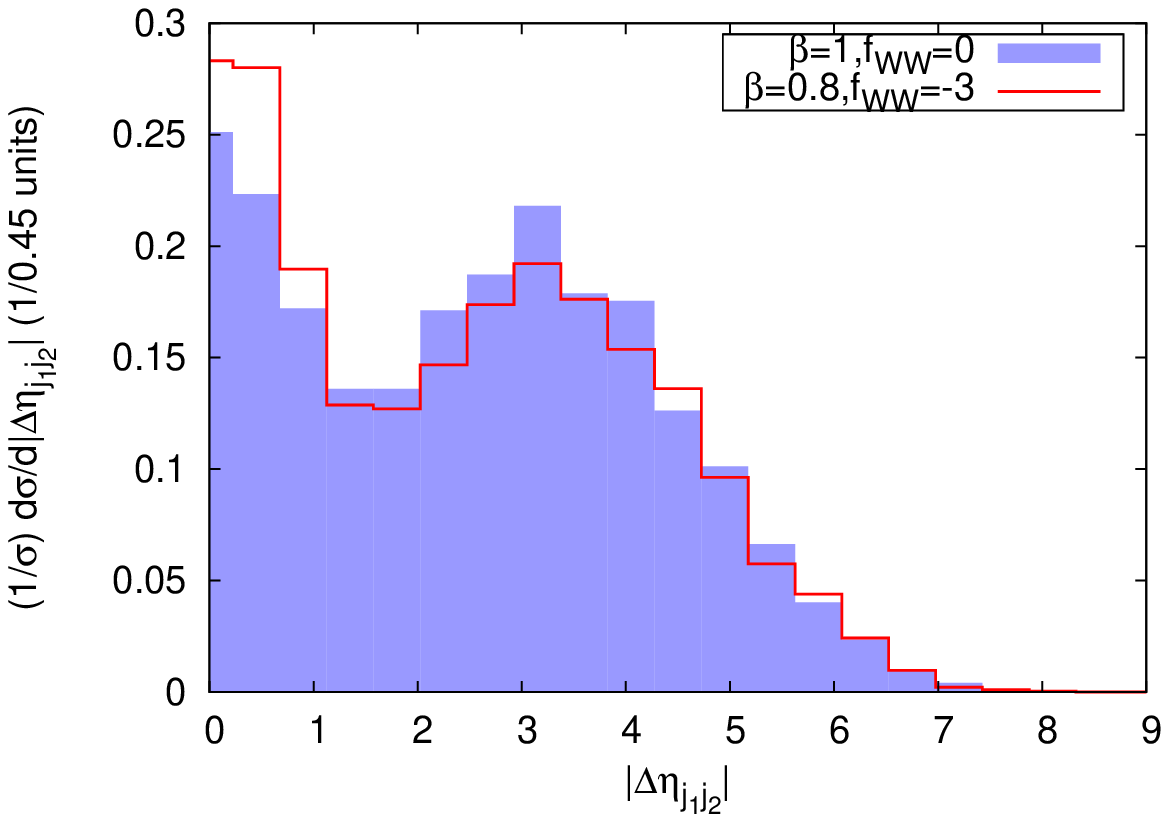}}
\centerline{\includegraphics[width=8.6cm,height=8.6cm,keepaspectratio]{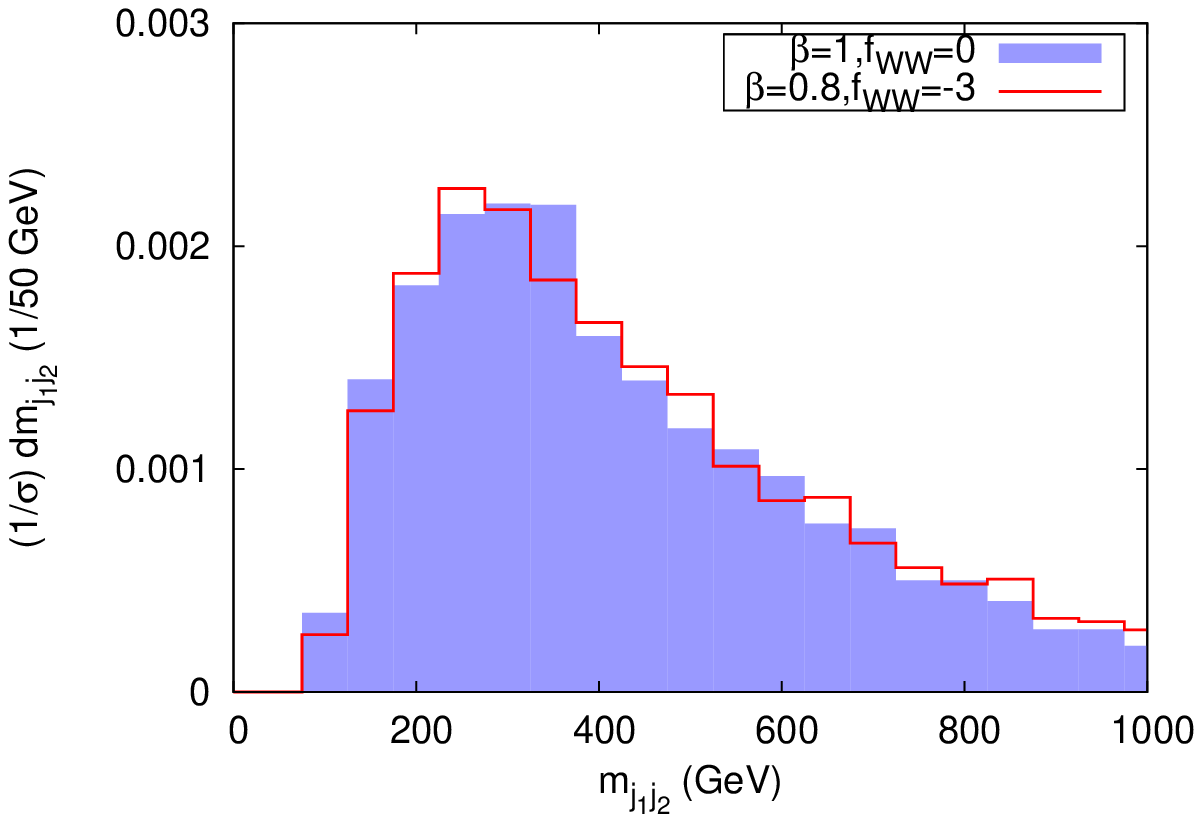}\
\includegraphics[width=8.6cm,height=8.6cm,keepaspectratio]{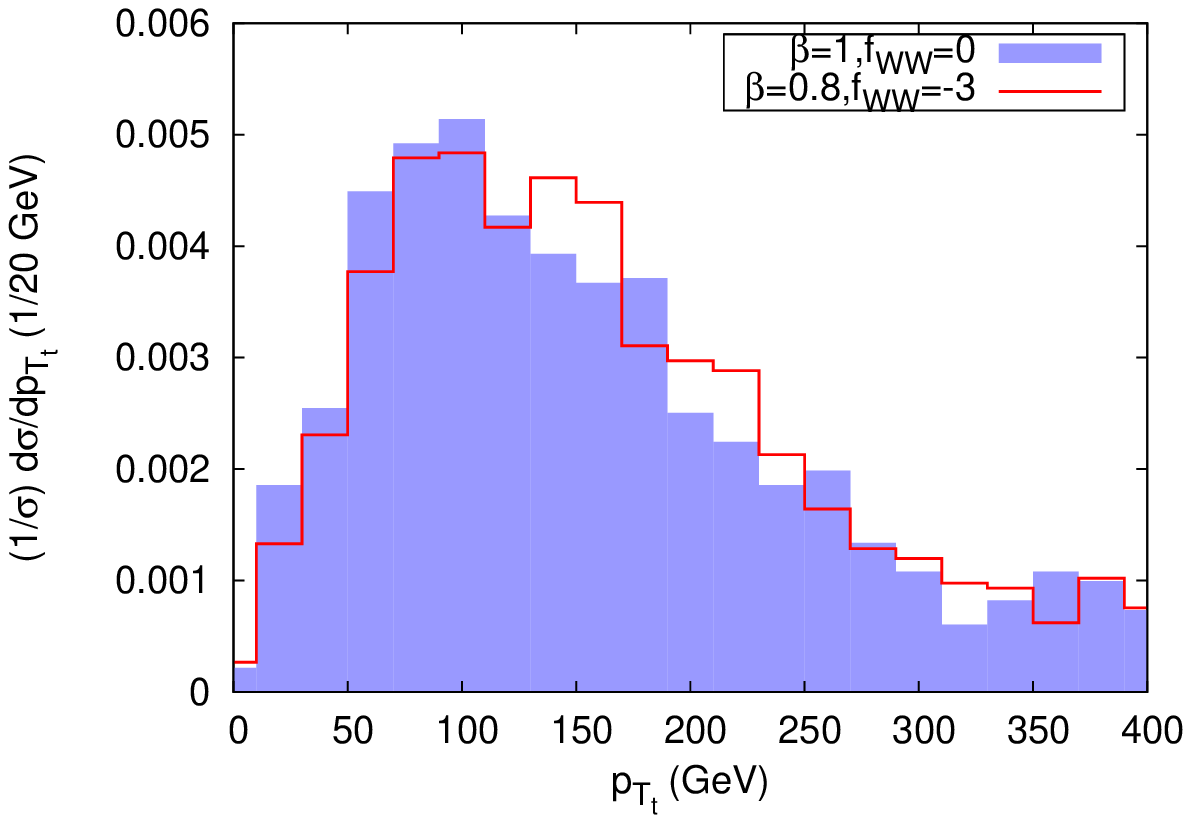}}
\centerline{\includegraphics[width=8.6cm,height=8.6cm,keepaspectratio]{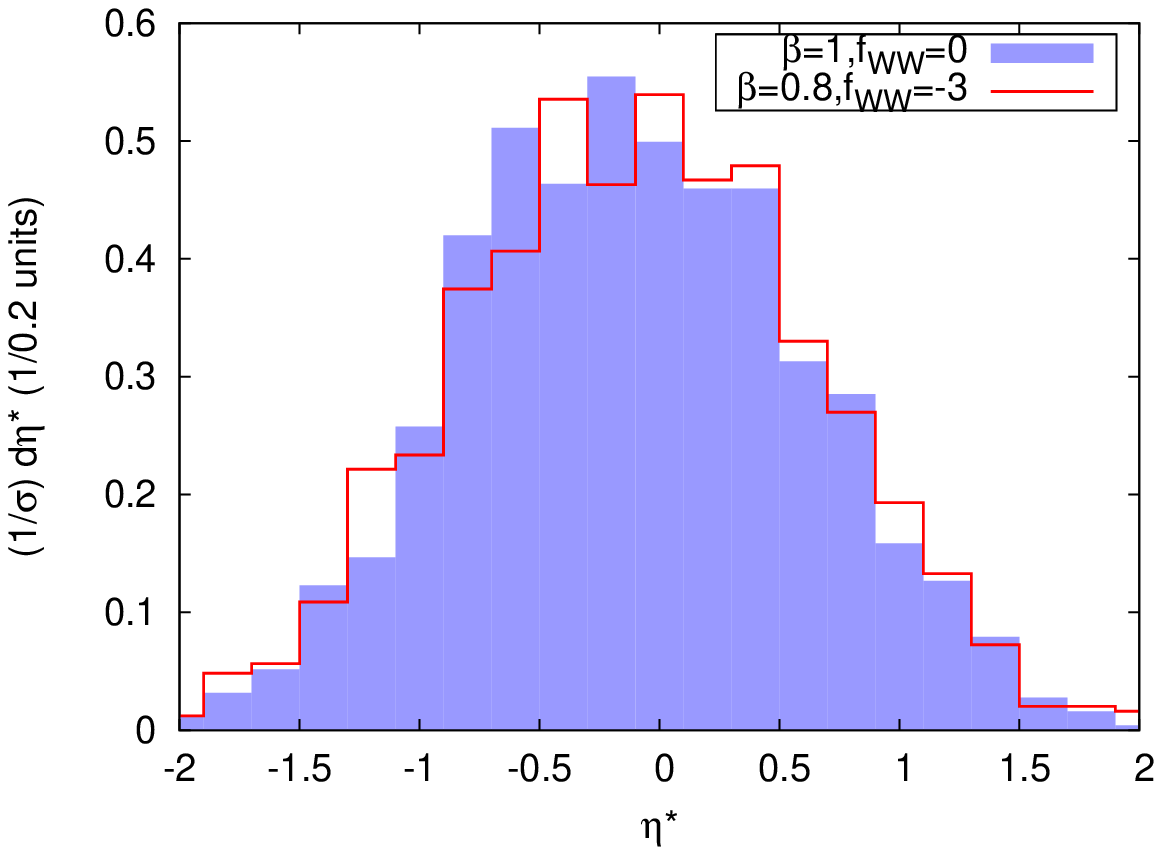}\
\includegraphics[width=8.6cm,height=8.6cm,keepaspectratio]{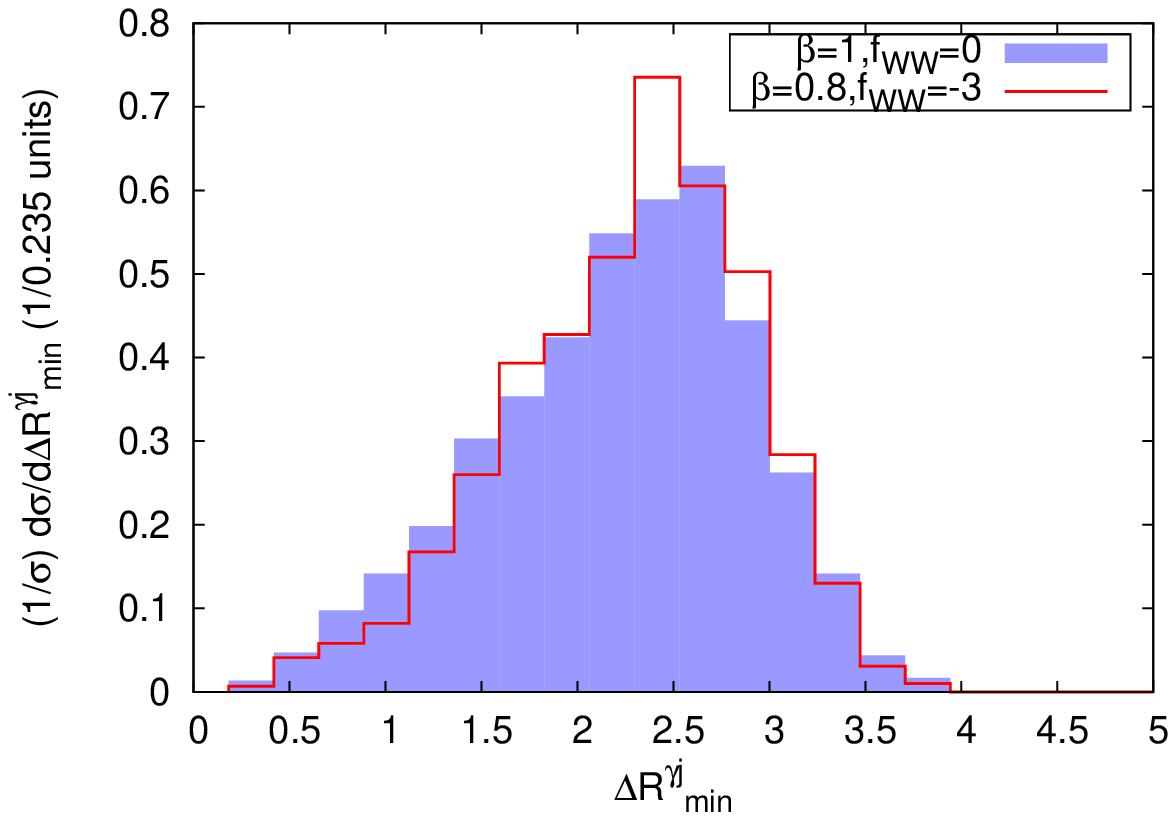}}
\caption{\small \sl Normalised distributions in various kinematic variables for 8 TeV LHC, top row : $\sqrt{\vec{p}_{Tj_{1}} \cdot \vec{p}_{Tj_{2}}}$ (left) and 
                    $|\Delta \eta_{j_{1} j_{2}}|$ (right); middle row : $m_{j_{1} j_{2}}$  (left) and $p_{T_{t}}$ (right); bottom row : $\eta^{*}$ (left) and 
                    $\Delta R_{\rm min}^{\gamma j}$ (right), for the parameter points $\{\beta = 1,f_{WW} = 0\}$ (shaded 
                    blue region) and $\{\beta = 0.8,f_{WW} = -3\}$ (solid red line). The cut-off scale chosen is $\Lambda=1$ TeV. Both the parameter points are allowed by the current data at 95\% C.L.}
\label{fig:dist-b1pt8-f0p3}
\end{figure} \\
\end{itemize}
 The last five kinematic variables form a subset of the inputs for the Boosted Decision Tree (BDT) employed by ATLAS for studying this channel~\cite{atlas3,atlas-gaga}.


Figure~\ref{fig:dist-b1-f0pm10} shows the normalised distributions in the above six variables for $\beta=1$ and $f_{WW}= 0,\pm10$ at the $8$ TeV LHC. The values $f_{WW}=\pm10$ are chosen for illustrative purpose only, 
since, as seen in section~\ref{sec-5}, although they are allowed by the LHC data in the $WW^{*}$ channel, the current measurement in the $\gamma \gamma$ channel restricts $f_{WW}$ to smaller values. 
Therefore, in figure~\ref{fig:dist-b1pt8-f0p3} we show the aforementioned distributions in the SM, and for the parameters 
$\{\beta,f_{WW}\} = \{0.8,-3\}$ the latter being within the $2\sigma$ allowed range of the global fit (see figure~\ref{margin}). For both 
the above figures, the cut-off scale has been chosen as $\Lambda=1$ TeV. We note that the distributions of $|\Delta \eta_{j_{1} j_{2}}|$ have two peaks. The peak at 
$|\Delta \eta_{j_{1} j_{2}}|=0$ is due to VH contamination. Moreover, the relative heights of the two peaks change on introducing higher dimensional operators.

\section{Summary}
\label{sec-8}
We have considered some illustrative dimension-6 operators for $HVV$ interactions, and their potential contributions to the Higgs data,
in conjunction with the SM-like operators. Parametrizing the strength of the additional interactions by $f$ ($f_{WW}/f_{BB}$), and the simultaneous
modification to the SM-like couplings to $W$ and $Z$ bosons by $\beta$, we show, after a detailed cut-based Monte Carlo analysis, how the efficiencies of different
acceptance cuts are altered for various values of $f$ and $\beta$. We find that in general there can be substantial modification of this
kind, which underscores the importance of a detailed study of the effect of all such additional operators on the kinematics of various
final states. When one further imposes the constraints on the ($f, \beta$) space as resulting from a global fit of the LHC data available
till date, the $f$-parameters are in general restricted to rather modest values while $\beta$ is restricted to be in the range $[0.8,1]$. Thus the effects of cuts 
in the diboson channels may not be drastically different, unless there is ground for relaxing their constraints. In general, the $VBF$ channel is more sensitive to the 
higher-dimensional operators than the gluon-fusion and $VH$ production modes. We also present several kinematic distributions, some of which are used in BDT analyses, which can 
potentially bring out signatures of the new operators, even with moderate strength.        

It should be remembered here that our analysis is purely phenomenological and data-driven; the assumption of any specific 
ultraviolet (UV) completion is deliberately avoided. In a specific UV completion scheme, more than one higher-dimensional operator 
relevant at the LHC scale may be generated, which can affect some of our conclusions. For example, with additional operators present, a
situation as restrictive as indicated by figure~\ref{fig:gaga-ww-total} may not arise due to the accidental cancellation of different contributions. However, 
studying one operator at a time gives us valuable insight on how it typically affects various observables in the Higgs 
sector --- an insight that is lost in the introduction of all operators of a given dimension simultaneously. In this spirit, we have 
explored two operators which can most significantly modify the interaction of the Higgs with a pair of electroweak gauge bosons.

\section*{Acknowledgment}
We thank Atri Bhattacharya, Sanjoy Biswas, Ushoshi Maitra and Tanumoy Mandal for discussions and technical help. Useful comments from Bruce Mellado are also acknowledged. The work of SM 
is supported by World Premier International Research Center Initiative (WPI Initiative), MEXT, Japan. SB and BM acknowledge the  funding available from the Department of Atomic Energy, 
Government of India, for the Regional Centre for Accelerator-based Particle Physics, Harish-Chandra Research Institute. Computational work for this study was partially carried out at the cluster computing facility at the Harish-Chandra Research Institute (http://cluster.hri.res.in). 


\end{document}